\documentclass[12pt]{article}

\usepackage{amsmath, amssymb}
\usepackage{graphicx}% Include figure files
\usepackage{caption}
\usepackage{color}% Include colors for document elements
\usepackage{dcolumn}% Align table columns on decimal point
\usepackage{bm}% bold math
\usepackage{float}
\usepackage{hyperref}
\usepackage{apacite}
\usepackage{algorithm}

\usepackage[table]{xcolor} 
\usepackage{multirow}

\hypersetup{colorlinks = true, citecolor = black, urlcolor = blue}

\usepackage{tikz}
\usetikzlibrary{positioning}

\definecolor{background-color}{gray}{0.98}
\definecolor{charlesBlue}{RGB}{100, 155, 255}

\title{A Review of Automatic Differentiation \\ and its Efficient Implementation}
\author{Charles C. Margossian\thanks{Department of Statistics, Columbia University -- contact: \textit{charles.margossian@columbia.edu}}}

\date{}

\begin{document}
\maketitle

%\begin{center}
%\subsubsection*{\small Article Type:}
%Overview

\hfill \break
\thanks

%\subsubsection*{Abstract}
%\begin{flushleft}
  %
  \begin{abstract}
  Derivatives play a critical role in computational statistics, examples being Bayesian inference
  using Hamiltonian Monte Carlo sampling and the training of neural networks.
  Automatic differentiation is a powerful tool to automate the calculation of derivatives
  and is preferable to more traditional methods, especially when differentiating complex algorithms and mathematical functions.
  The implementation of automatic differentiation however requires some care to insure efficiency.
  Modern differentiation packages deploy a broad range of computational techniques to improve applicability, run time,
  and memory management.
  Among these techniques are operation overloading, region-based memory, and expression templates.
  There also exist several mathematical techniques which can yield high performance gains when applied to complex algorithms.
  For example, semi-analytical derivatives can reduce by orders of magnitude the runtime required to numerically solve and differentiate an algebraic equation.
  Open and practical problems include the extension of current packages to provide more specialized routines,
  and finding optimal methods to perform higher-order differentiation.
  \end{abstract}
  %   
%\end{flushleft}
%\end{center}
%
%\clearpage
%
%\renewcommand{\baselinestretch}{1.5}
\normalsize

\clearpage

\section*{Graphical table of content}

  \begin{figure}[htbp]
  \begin{center}
  \begin{tikzpicture}
  [
    Box/.style={rectangle, draw=black!, fill=green!0, thick, minimum size=10mm},
    Gray/.style={rectangle, draw=black!, fill=gray!35, thick, minimum size=10mm}
  ]
  % Nodes
  \node[Gray] (AD) at(0, 0) {\ \ \large \textbf{Automatic Differentiation} \ \ };
  
  \node[Box, text width = 6cm] (goal) at(10, 0) {\noindent \begin{center} Differentiate math \\ functions and algorithms \end{center} \ \ };
  
  \node[Box, text width = 8cm] (imp) at (0, -5) 
     {\noindent 
      \textbf{Computational implementation}
      \begin{itemize} \vspace{-2.5mm}
      \itemsep-0.5em 
        \item Source transformation
        \item Operator overloading
        \item Checkpointing
        \item Region-based memory
        \item Expression templates
       \end{itemize}
       
       \noindent
       \textbf{Mathematical implementation}
       \begin{itemize} \vspace{-2.5mm}
       \itemsep-0.5em
         \item Forward and reverse-mode chain rule
         \item Reduced expression graph
         \item Super nodes
       \end{itemize}
       \ \\
     };
     
  \node[Box, text width = 8cm] (app) at (10, -5) {
    \textbf{Applications} %\vspace{-2.5mm}
    \begin{itemize}
       \item Hamiltonian Monte Carlo sampling
       \item Variational inference
       \item Neural networks
       \item Numerical opitmization 
    \end{itemize}
    \ \\
  };
                            
  % Lines
   \path [->, draw, thick] (imp) -- (AD);
   \path [->, draw, thick] (AD) -- (goal);
   \path [->, draw, thick] (goal) -- (app);
  
  \end{tikzpicture}
  \end{center}
  \label{fig:Expressiongraph}
  \caption*{\textit{Automatic differentiation is a powerful tool to differentiate mathematical functions and algorithms. 
  It has, over the past years, been applied to many branches of computational statistics. This article reviews some important 
  mathematical and computational considerations required for its efficient implementation.}}
  \end{figure}
  
\section*{\sffamily \Large Introduction}

  A large number of numerical methods require derivatives to be calculated.
  Classical examples include the maximization of an objective function using gradient descent or Newton's method.
  In machine learning, gradients play a crucial role in the training of neural networks \cite{Widrow:1990}.
  Modelers have also applied derivatives to sample the high-dimensional posteriors of Bayesian models, 
  using Hamiltonian Monte 
  Carlo samplers \cite{Radford:2010, Betancourt:2017} or approximate these posteriors with variational inference \cite{Kucukelbir:2016}.
  As probabilistic models and algorithms gain in complexity, computing derivatives becomes a formidable task. 
  For many problems, we cannot calculate derivatives analytically, but only evaluate them numerically at certain points.
  Even when an analytical solution exists, working it out by hand can be mathematically challenging, time consuming, and error prone.
  
  There exist three alternatives that automate the calculation of derivatives: 
  (1) finite differentiation, (2) symbolic differentiation, and (3) automatic differentiation (AD). 
  The first two methods can perform very poorly when applied to complex functions for a variety of reasons summarized in Table~\ref{tab:techniques}; 
  AD on the other hand, escapes many limitations posed by finite and symbolic differentiation, as discussed by \cite{BaydinEtAl:2015}.
  Several packages across different programing languages implement AD and
  allow differentiation to occur seamlessly, while users focus on other programing tasks.
  For an extensive and regularly updated list of AD tools, the reader may consult \url{www.autodiff.org}. 
  In certain cases, AD libraries are implemented as black boxes which support statistical and machine learning softwares,
  such as the python package \texttt{PyTorch} \cite{Paszke:2017} or the probabilistic programing language \texttt{Stan} \cite{Stan:2017}.
  % Table~\ref{tab:ADsoftwares} provides an overview of the packages
  % we explicitly mention in this paper.  %, but this list is by no means comprehensive.

  Automatic differentiation does not, despite its name, fully automate differentiation and can yield inefficient code if naively implemented.
  For this reason, AD is sometimes referred to as \textit{algorithmic} rather than \textit{automatic} differentiation.
  % (see for example \cite{Griewank:2003}).
  We here review some important techniques to efficiently implement AD and optimize its performance.
  The article begins with a review of the fundamentals of AD.
  The next section presents various computational schemes, required for an efficient implementation and deployed in several recently developed packages.
  We then discuss strategies to optimize the differentiation of complex mathematical algorithms.
  The last section presents practical and open problems in the field.
  
  Throughout the paper, we consider several performance metrics. 
  Accuracy is, we argue, the most important one, but rarely distinguishes different implementation schemes of AD,
  most of which are exact up to arithmetic precision.
  On the other hand, differences arise when we compare run time and memory usage;
  run time, being easier to measure, is much more prominent in the literature.
  Another key consideration is applicability: does a method solve a broad or only a narrow class of problems?
  Finally, we discuss ease of implementation and readability, more subjective but nevertheless essential properties of a computer code.
  We do not consider compile time, mostly because the statistical applications of AD we have in mind compile a program once,
  before using it thousands, sometimes millions of times.
  
  \setlength{\extrarowheight}{5pt}
  \begin{table}
  \scriptsize
  \begin{center}
  \begin{tabular}{l l l}
  \rowcolor[gray]{0.95}  \textbf{Technique} & \textbf{Advantage(s)} & \textbf{Drawback(s)} \\ %& Complexity \\
  %\hline \hline
   %\hline
  \textbf{Hand-coded analytical} & Exact and often fastest &  Time consuming to code, error prone, and not\\ %& $O(1)$ \\
  \textbf{derivative} & method. & applicable to problems with implicit solutions. \\
       & & Not automated. \vspace{2mm} \\
  %\hline
  \rowcolor[gray]{0.95}  \textbf{Finite differentiation} & Easy to code. & Subject to floating point precision errors and slow, \\
  \rowcolor[gray]{0.95}  & & especially in high dimensions, as the method requires \\ 
  \rowcolor[gray]{0.95}  & &  at least $D$ evaluations, where $D$ is the number of \\
   \rowcolor[gray]{0.95} & &   partial derivatives required. \vspace{2mm} \\
  % \hline
  \textbf{Symbolic differentiation} & Exact, generates symbolic & Memory intensive and slow.  Cannot handle\\
     & expressions. & statements such as  unbounded loops.\vspace{2mm} \\ 
   % \hline
   \rowcolor[gray]{0.95} \textbf{Automatic differentation} & Exact, speed is comparable & Needs to be carefully implemented, alhough this \\
   \rowcolor[gray]{0.95}  & to hand-coding derivatives, & is already done in several packages. \\
   \rowcolor[gray]{0.95} & highly applicable. & \\
   % \hline
  \end{tabular}
  \caption{Summary of techniques to calculate derivatives.}
  \label{tab:techniques}
  \end{center}
  \end{table}
  
%  \begin{table}
%  \scriptsize
%  \begin{center}
%  \begin{tabular}{l l l l}
%  Language & Package & Reference & URL \\
%  \hline \hline
%  \\
%  C++ & CasADi & \cite{Andersson:2018} & \url{https://github.com/casadi/casadi} \\
%   & & & (Note: CasADi is also available in Python) \\
%   & Stan Math & \cite{Carpenter:2015} & \url{mc-stan.org} \\
%   & Adept & \cite{Hogan:2014} & \url{http://www.met.reading.ac.uk/clouds/adept/} \\
%   & CppAD & \cite{Bell:2012} & \url{http://www.coin-or.org/CppAD} \\
%   & ADMB & \cite{Fournier:2012} & \url{http://www.admb-project.org/} \\
%   & Adol-C & \cite{Griewank:2008} & \url{https://projects.coin-or.org/ADOL-C} \\
%   & Sacado & \cite{Gay:2005} & \url{http://trilinos.org} \\
%%  \\
%%   C\# & AutoDiff & \cite{Shtof:2013} & \url{http://autodiff.codeplex.com/} \\
%   \\
%   Fortran & ADIFOR & \cite{Bischof:1996} & \url{http://www.mcs.anl.gov/research/ projects/adifor/} \\
%%   \\
%%   Matlab & ADiMat & \cite{Willkomm:2013} & \url{http://adimat.sc.informatik.tu- darmstadt.de/} \\
%   \\
%   Python & PyTorch & \cite{Paszke:2017} & \url{https://pytorch.org/} \\
%%               & ad & & \url{https://pypi.python.org/pypi/ad} \\
%%               & autograd &   & \url{https://github.com/HIPS/autograd} \\
%%   \\
%%   R & Madness & \cite{Pav:2016} & \url{https://github.com/shabbychef/madness} \\
%%     \\
%%   Julia & JuliaDiff & & \url{http://www.juliadiff.org/} \\
%  \hline
%   \end{tabular}
%  \caption{Overview of packages for automatic differentiation.
%  \textit{This table contains the packages explicitly mentioned in this paper. 
%  For a more comprehensive and regularly updated list, the user may consult \url{www.autodiff.org}.
%  The packages are ordered by languages and by the publication date of the corresponding reference.}}
%  \label{tab:ADsoftwares}
%  \end{center}
%  \end{table}

  \section*{\sffamily \Large How automatic differentiation works} \label{Mechanism}
 
 Given a target function $f:\mathbb R^n \to \mathbb R^m$,
 the corresponding $m \times n$ Jacobian matrix $J$ has $(i, j)^\mathrm{th}$ component:
  $$ J_{ij} = \frac{\partial f_i}{\partial x_j} $$
  This matrix contains the partial derivatives of all the outputs with respect to all
  the inputs. If $f$ has a one-dimensional output, as is the case when $f$ is an objective function,
  the Jacobian matrix is simply the gradient. 
  In practice we may care only about the partial derivatives with respect to some of the inputs and 
  calculate a reduced Jacobian matrix.
  The partial derivatives with respect to these inputs, 
  respectively the corresponding columns of the Jacobian matrix, are called \textit{sensitivities}.
%  Often we compute the sensitivities for parameters, as we explore a parameter space, 
%  but not for data, which stay fixed.
  
  Now suppose $f$ is a composite function: $f(x) = h \circ g (x) = h(g(x))$, 
  with $x \in \mathbb R^n$, \mbox{$g: \mathbb R^n \to \mathbb R^k$} and \mbox{$h:\mathbb R^k \to \mathbb R^m$}. 
  Applying the chain rule and elementary matrix multiplication:
  $$ J = J_{h \circ g} = J_h(g(x)) \cdot J_g(x) $$ 
  with $(i, j)^\mathrm{th}$ element:
  $$ J_{ij} = \frac{\partial f_i}{\partial x_j} = \frac{\partial h_i}{\partial g_1}  \frac{\partial g_1}{\partial x_j}
  +  \frac{\partial h_i}{\partial g_2}  \frac{\partial g_2}{\partial x_j} + ... +
  \frac{\partial h_i}{\partial g_k}  \frac{\partial g_k}{\partial x_j} $$
  More generally, if our target function $f$ is the composite expression of $L$ functions,
  \begin{eqnarray}
    f = f^L \circ f^{L - 1} \circ ... \circ f^1
    \label{eq:compFunction}
  \end{eqnarray}
  the corresponding Jacobian matrix verifies:
  \begin{eqnarray}
    J = J_L \cdot J_{L-1} \cdot ... \cdot J_1
    \label{eq:chainJacobian}
  \end{eqnarray}

  To apply AD, we first need to express our target function $f$ using a computer program.
  AD can then operate in one of two modes: \textit{forward} or \textit{reverse}.
  Let $u \in \mathbb R^n$.
  One application or \textit{sweep} of forward-mode AD \underline{numerically} evaluates the action of the Jacobian matrix on $u$,
  $J \cdot u$. As prescribed by Equation~\ref{eq:chainJacobian},
  \begin{eqnarray}
    \begin{aligned}
    J \cdot u & = J_L \cdot J_{L - 1} \cdot ... \cdot J_3 \cdot J_2 \cdot J_1 \cdot u \\
                   & = J_L \cdot J_{L - 1} \cdot ... \cdot J_3 \cdot J_2 \cdot u_1 \\
                   & = J_L \cdot J_{L - 1} \cdot ... \cdot J_3 \cdot u_2 \\
                   & . . . \\
                   & = J_L \cdot u_{L - 1} \\
    \end{aligned}
  \end{eqnarray}
  where the $u_l$'s verify the recursion relationship
  \begin{eqnarray}
    \begin{aligned}
    u_1 & = J_1 \cdot u \\
    u_l & = J_l \cdot u_{l - 1} \\
    \end{aligned}
    \label{eq:recursion}
  \end{eqnarray}
  Hence, given a complex function $f$, we can break down the action of the Jacobian matrix on a vector,
  into simple components, which we evaluate sequentially.
  Even better, we can choose splitting points in equation~\ref{eq:compFunction} (correspondingly equation~\ref{eq:chainJacobian}) 
  to generate efficient implementations of AD.
  We will take advantage of this fact when we consider checkpoints and super nodes in later sections of the paper.
  
  Consider now a vector in the output space, $w \in \mathbb R^m$. A sweep of reverse-mode AD computes the
  action of the transpose of the Jacobian matrix on $w$, $J^T \cdot w$. The reasoning we used for forward mode applies here too
  and allows us to break down this operation into a sequence of simple operations.
  Often times, an AD program evaluates these simple operations analytically.
  
%  To meet the above criterion, a function of interest is often broken into elementary operators, such as \texttt{+} or \texttt{log} (addition or logarithm).
%  Hence, an intermediate function $f^l:\mathbb R^{l_\mathrm{in}} \to \mathbb R^{l_\mathrm{out}}$
%  may only alter one or two of its input elements, and apply an \textit{identity} operation to the other elements.
%  By linearity of equation~\ref{eq:recursion}, the identity is preserved when we compute $u_l$.  
%  This means that, for $l \neq k$, $u_l$ and $u_k$ may share several elements, due to the linearity of equation~\ref{eq:recursion}.
  
  To further illustrate the mechanisms of AD, we consider a typical example from 
  statistics\footnote{Based on an example from \cite{Carpenter:2015}.}.
  We will compute the gradient of a log likelihood function,
  for an observed variable $y$ sampled from a normal distribution.
  The likelihood function is:
  \begin{eqnarray*}
    \mathrm{Normal}(y \mid \mu, \sigma^2) = \frac{1}{\sqrt{2 \pi} \ \sigma} \exp \left(- \frac{1}{2 \sigma^2} (y - \mu)^2 \right)
  \end{eqnarray*}
  with corresponding log likelihood function:
  \begin{eqnarray}
    f(y, \mu, \sigma) = \log(\mathrm{Normal}(y \mid \mu, \sigma^2)) 
                               = -\frac{1}{2} \left(\frac{y - \mu}{\sigma} \right)^2 - \log(\sigma) - \frac{1}{2}\log(2\pi)
  \label{eq:logNormal}
  \end{eqnarray}
  Our goal is to compute the sensitivities of $\mu$ and $\sigma$, evaluated at $y = 10$, $\mu = 5$, and $\sigma = 2$.
  The above function can be broken up into a sequence of maps, 
  yielding the composite structure of Equation~\ref{eq:compFunction}:
  \begin{eqnarray}
    \begin{aligned}
    (y, \ \mu, \ \sigma) \to & (y - \mu, \ \sigma) \\
                              \to & \left(\frac{y - \mu}{\sigma}, \ \sigma \right) \\
                              \to & \left(\frac{y - \mu}{\sigma}, \ \log(\sigma) \right) \\
                              . . . & \\
                              \to & -\frac{1}{2} \left(\frac{y - \mu}{\sigma} \right)^2 - \log(\sigma) - \frac{1}{2}\log(2\pi)
    \end{aligned}
    \label{eq:maps}
  \end{eqnarray}
  Equation~\ref{eq:recursion}, which gives us the sequential actions of Jacobian matrices on vectors, then applies.
  Note that we have broken $f$ into functions which only perform an elementary operation, such as \texttt{/} or \texttt{log}
  (division or logarithm). For example, the second map is $f^2:(\alpha, \beta) \to (\alpha / \beta, \beta)$,
  and constructs a two dimensional vector by applying the \textit{division} and the \textit{identity} operators.
  By linearity of Equation~\ref{eq:recursion}, the identity operation is preserved.
  This means that, in our example, the second elements of $u_1$ and $u_2$ are identical.
  Hence, rather than explicitly computing every element in Equation~\ref{eq:recursion},
  i.e. propagating derivatives through the maps outlined in Equation~\ref{eq:maps},
  it suffices to focus on individual operations, such as \texttt{/} and \texttt{log},
  and moreover, any operation other than the identity operation.
  
  To do this, Equation~\ref{eq:logNormal} can be topologically sorted into an
  \textit{expression graph} (Figure~\ref{fig:Expressiongraph}). At the top of the 
  graph, we have the final output,  $f(y, \mu, \sigma)$, and at the roots the 
  input variables $y$, $\mu$, and $\sigma$. The nodes in between represent intermediate 
  variables, obtained through elementary operations. A link between variables on the graph indicates an explicit dependence.
  We then know how to analytically differentiate operators on the expression graph and can get
  partial derivatives of composite expressions with the chain rule. For example:
  $$
    \frac{\partial v_5}{\partial \mu} = \frac{\partial v_5}{\partial v_4} \times \frac{\partial v_4} {\partial \mu}
                                                    = \frac{1}{\sigma} \times (-1)
  $$
  
  We now describe the algorithmic mechanisms of the forward and reverse modes of AD.
  
  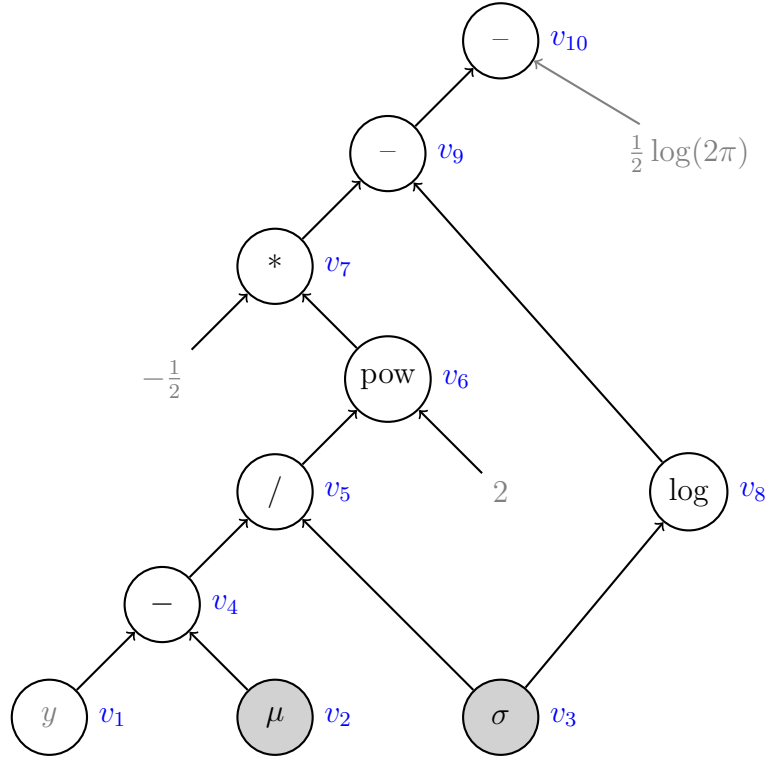
\begin{figure}
  \begin{center}
  \begin{tikzpicture}
  [
    Round/.style={circle, draw=black!, fill=green!0, thick, minimum size=10mm},
    Red/.style={circle, draw=black!, fill=red!255, thick, minimum size=10mm},
    Yellow/.style={circle, draw=black!, fill=yellow!255, thick, minimum size=10mm},
    Gray/.style={circle, draw=black!, fill=gray!35, thick, minimum size=10mm}
  ]
  % Nodes
  \node[Round, label=right:{\textcolor{blue}{$v_{10}$}}] (v10) at(0, 0) {--};
  
  \node[Round, label=right:{\textcolor{blue}{$v_{9}$}}] (v9) at(-1.5, -1.5) {--};
  \node (constant1) at(2.5, -1.5) {\textcolor{gray}{$\frac{1}{2} \log(2 \pi)$}};
  
  \node[Round, label=right:{\textcolor{blue}{$v_7$}}] (v7) at(-3, -3) {*};
  
  \node (constant2) at(-4.5, -4.5) {\textcolor{gray}{$- \frac{1}{2}$}};
  \node[Round, label=right:{\textcolor{blue}{$v_6$}}] (v6) at(-1.5, -4.5) {pow};
  
  \node[Round, label=right:{\textcolor{blue}{$v_5$}}] (v5) at(-3, -6) {/};
  \node (constant3) at(0, -6) {\textcolor{gray}{$2$}};
  \node[Round, label=right:{\textcolor{blue}{$v_8$}}] (v8) at(2.5, -6) {log};
  
  \node[Round, label=right:{\textcolor{blue}{$v_4$}}] (v4) at(-4.5, -7.5) {$-$};
  
  \node[Round, label=right:{\textcolor{blue}{$v_1$}}] (v1) at(-6, -9) {\textcolor{gray}{$y$}};
  \node[Gray, label=right:{\textcolor{blue}{$v_2$}}] (v2) at(-3, -9) {$\mu$};
  \node[Gray, label=right:{\textcolor{blue}{$v_3$}}] (v3) at(0, -9) {$\sigma$};

  % Lines
  \path [<-, draw, thick] (v10) -- (v9);
  \path [<-, draw = gray, thick] (v10) -- (constant1);
  \path [<-, draw, thick] (v9) -- (v7);
  \path [<-, draw, thick] (v7) -- (constant2);
  \path [<-, draw, thick] (v7) -- (v6);
  \path [<-, draw, thick] (v6) -- (v5);
  \path [<-, draw, thick] (v6) -- (constant3);
  \path [<-, draw, thick] (v9) -- (v8);
  \path [<-, draw, thick] (v5) -- (v4);
  \path [<-, draw, thick] (v4) -- (v1);
  \path [<-, draw, thick] (v4) -- (v2);
  \path [<-, draw, thick] (v5) -- (v3);
  \path [<-, draw, thick] (v8) -- (v3);
  
  \end{tikzpicture}
  \end{center}
  \caption{Expression graph for the log-normal density. \textit{The above graph is generated by
  the computer code for Equation~\ref{eq:logNormal}. Each node represents a variable, labeled $v_1$
  through $v_{10}$, which is calculated by applying a mathematical operator to variables lower
  on the expression graph. 
  The top node ($v_{10}$) is the output variable
  and the gray nodes the input variables for which we require sensitivities. 
  The arrows represent the flow of information when
  we numerically evaluate the log-normal density. Adapted from Figure 1 of \protect\cite{Carpenter:2015}.}}
  \label{fig:Expressiongraph}
  \end{figure}
  
  \subsection*{\sffamily \large Forward mode}
  
  % AD proceeds in several \textit{sweeps}. 
  % Recall our target function is $f(x)$ with $x \in \mathbb R^n$ and $J$ is its matrix .
  Consider the direction, or \textit{initial tangent}, $u \in \mathbb R^n$.
  One forward sweep computes $J \cdot u$.
  This corresponds to the total derivative of $f$ with respect to $u$, or to be more precise
  to the directional or directed partial derivative.
  The complexity of this operation is linear in the complexity of $f$.
  Using fused-multiply adds, denoted OPS, as  a metric for computational
  complexity, 
   $ \mathrm{OPS}(f(x), J \cdot u(x)) \le 2.5  \times \mathrm{OPS}(f(x))$ 
  (see chapter 4 of \cite{Griewank:2008}).

  The right choice of $u$ allows us to compute one column of the Jacobian matrix,
  that is the partial derivatives of all the outputs with respect to one input.
  Specifically, let $u_j$ be 1 for the $j^\mathrm{th}$ input and 0 for all the other inputs.
  Then:
  \begin{eqnarray}
    \begin{aligned}
    J_{.j} & = J \cdot u_j  \\
     \end{aligned}
  \label{eq:fwdSweep}
  \end{eqnarray}
  Here $J_{.j}$ denotes the $j^\mathrm{th}$ column of the matrix $J$.
  We can hence compute a full $m \times n$ Jacobian 
  matrix in $n$ forward sweeps. Naturally, we do not compute Equation~\ref{eq:fwdSweep} by doing
  a matrix operation, as this would require us to already know $J$. Instead, we proceed as follows.

  Forward mode AD computes a directional derivative at the same time as it performs a forward evaluation trace.
  % To be more precise, it computes, in a given coordinate system, a directed partial derivative.
  %
  In the case where $u_j$ is the above defined initial tangent, the software evaluates a variable and then
  calculates its partial directed derivative with respect to one root variable.
  % That is, after evaluating a variable, the software calculates its total derivative with respect to one root variable.
  Let $\dot v$ denote the directed partial derivative of $v$ with respect to that one root variable.
  Consider $v_5$, which is connected to $v_4$ and $v_3$. 
  As the program sweeps through $v_5$ it evaluates \mbox{(i) its value} and (ii) its directional derivative.
  The latter can be calculated by computing the partial derivatives of $v_5$ with
  respect to $v_4$ and $v_3$, and plugging in numerical values for $v_3$, $v_4$,
  and their directed partial derivatives $\dot v_3$ and $\dot v_4$, which have already been evaluated. 
  The code applies this procedure from the roots to the final output (Table~\ref{table:forwardAD}).

  \begin{table}  
    \centering
    \renewcommand{\arraystretch}{1.5}
    \begin{tabular}{l l l l l}
    \textbf{Forward evaluation trace} & \textbf{Forward derivative trace} \\
    % \hline
    \rowcolor[gray]{0.95}  $v_1 = y = 10$ & $\dot v_1 = 0$ \\
    \rowcolor[gray]{0.95}  $v_2 = \mu = 5$  & $\dot v_2 = 1$ \\
    \rowcolor[gray]{0.95}  $v_3 = \sigma = 2$ & $\dot v_3 = 0$ \\
    $v_4 = v_1 - v_2 = 5$ &  $\dot{v}_{4} = \frac{ \partial v_4}{\partial v_1} \dot v_1 + \frac{ \partial v_4}{\partial v_2 } \dot v_2 = 
      0 + (-1) \times 1 = -1$ \\  
    $v_5 = v_4 / v_3 = 2.5$ & $\dot v_5 = \frac{\partial v_5}{\partial v_4} \dot v_4 + \frac{\partial v_5}{\partial v_3} \dot v_3
                                                                  = \frac{1}{v_3} \times (-1) + 0 = - 0.5$ \\
    $v_6 = v_5^2 = 6.25$ & $\dot v_6 = \frac{\partial v_6}{\partial v_5} \dot v_5 = 2 v_5 \times (-0.5) =  -2.5$ \\
    $v_7 = -0.5 \times v_6 = 3.125$ & $\dot v_7 = \frac{\partial v_7}{\partial v_6} \dot v_6 = -0.5 \times (-2.5) = 1.25$ \\ 
    $v_8 = \log(v_3) = \log(2)$ & $\dot v_8 = \frac{\partial v_8}{\partial v_3} \dot v_3 = 0$ \\
    $v_9 = v_7 - v_8 = 3.125 - \log(2)$ & $\dot v_9 = \frac{\partial v_9}{\partial v_7} \dot v_7 + \frac{\partial v_9}{\partial v_8} \dot v_8 
    = 1 \times 1.25 + 0 = 1.25$\\
    \rowcolor[gray]{0.95}  $v_{10} = v_9 - 0.5 \log(2\pi) =3.125 - \log(4\pi) \ \ \ \ $ & $\dot v_{10} = \frac{\partial v_{10}}{\partial v_9} \dot v_9 = 1.25$ \\
    % \hline
    \end{tabular}
    \caption{Forward-mode AD.
                 \textit{
                  The forward derivative trace computes the derivative of a log Normal density with respect
                  to $\mu$. The program begins by initializing the derivatives of the root variables (1 for
                  $\mu$, 0 for the other inputs). The directional derivative is then computed at each node and
                  numerically evaluated using the chain rule, and previously evaluated variables and
                  directional derivatives. To get the sensitivity for $\sigma$, we must compute a new forward derivative trace.}}
    \label{table:forwardAD}
\end{table}

  \subsection*{\sffamily \large Reverse mode}

  Now suppose instead of computing derivatives with respect to an input, we
  compute the \textit{adjoints} with respect to an output. The adjoint of a variable $x$ with respect 
  to another variable $z$ is defined as:
   $$ \bar x = \frac{\partial z}{\partial x} $$
   Then for an initial \textit{cotangent} vector, $\bar u \in \mathbb R^m$, one reverse mode sweep computes
   $J^T \bar u$, with complexity $\mathrm{OPS}(f(x), J^T \bar u(x)) \le 4 \times \mathrm{OPS}(f(x))$
   (see again chapter 4 of \cite{Griewank:2008}).
   
   The right choice of $w$ allows us to compute one row of the Jacobian matrix, i.e. the adjoints of all the inputs
   with respect to one output. If the function has a one-dimensional output, this row corresponds exactly to the gradient.
   If we pick $w$ to be 1 for the $i^\mathrm{th}$ element and 0
   for all other elements, one sweep computes the $i^\mathrm{th}$ row of $J$:
   \begin{eqnarray}
     J_{i.} = J^T \bar w_i
   \label{eq:reverse}
   \end{eqnarray}
   and we get the full $m \times n$ Jacobian matrix in $m$ reverse sweeps.

   To calculate the right-hand side of Equation~\ref{eq:reverse}, the algorithm proceeds as follows. 
   After executing a forward evaluation trace, 
   as was done for forward mode, the program makes a reverse pass to calculate the adjoints. We
    start with the final output, setting its adjoint with respect to itself to 1, and compute successive adjoints until we
   reach the root variables of interest (Table~\ref{table:reverseMode}). 
   
   This procedure means the program must go through the expression graph twice: one 
   forward trace to get the value of the function and the intermediate variables;
   and one reverse trace to get the gradient. This creates performance overhead because the 
   expression graph and the values of the intermediate variables need to be stored in memory.
   One way of doing this efficiently is to have the code make a 
   \textit{lazy evaluation} of the derivatives. A lazy evaluation does \textit{not} evaluate a statement immediately
   when it appears in the code, but stores it in memory and evaluates it when (and only if) the
   code explicitly requires it. As a result, we only evaluate expressions required to compute the gradient,
   or more generally the object of interest.
   There are several other strategies to efficiently handle the memory overhead of reverse-mode AD, 
   which we will discuss in the section on \textit{computational implementation}.
%   This scheme is implemented in several packages for AD, notably
%   \texttt{Stan Math} \cite{Carpenter:2015} and several other packages we will discuss when we review template expressions in section.
   
    \begin{table}  
    \centering
    \renewcommand{\arraystretch}{1.5}
    \begin{tabular}{l l l}
    \textbf{Reverse adjoint trace} \\
    %\hline
    \rowcolor[gray]{0.95} $\bar v_{10} = 1$ \\
    $\bar v_9 = \frac{\partial v_{10}}{\partial v_9} \bar v_{10} = 1 \times 1 = 1$ \\
    $\bar v_8 = \frac{\partial v_9}{\partial v_8} \bar v_9  = (-1) \times 1 = -1$ \\
    $\bar v_7 = \frac{\partial v_9}{\partial v_7} \bar v_9 = 1 \times 1 = 1$ \\
    $\bar v_6 = \frac{\partial v_7}{\partial v_6} \bar v_7 = (-0.5) \times 1 = -0.5$ \\
    $\bar v_5 = \frac{\partial v_6}{\partial v_5} \bar v_6 = 2 v_5 \times \bar v_6 = 2 \times 2.5 \times (-0.5) = -2.5$ \\
    $\bar v_4 = \frac{\partial v_5}{\partial v_4} \bar v_5 = \frac{1}{v_3} \times (-2.5) = 0.5 \times (-2.5) = -1.25$ \\
    %\hline
    \rowcolor[gray]{0.95} $\bar v_3 = \frac{\partial v_5}{\partial v_3} \bar v_5 + \frac{\partial v_8}{\partial v_3} \bar v_8 
                    = - \frac{v_4}{v_3^2} \times (-2.5) + \frac{1}{v_3} \times (-1) = 2.625$ \\
    \rowcolor[gray]{0.95} $\bar v_2 = \frac{\partial v_4}{\partial v_2} \dot v_4 = (-1) \times (-1.25) = 1.25$ \\
    \end{tabular}
    \caption{Reverse-mode AD.
                  \textit{After executing a forward evaluation trace,
                  the reverse derivative trace computes the gradient of the log Normal density with respect to $\mu$ and $\sigma$.
                  For each node, an adjoint is computed and then combined with previously calculated adjoints
                  using the chain rule. This process should be compared to forward mode AD, depicted in 
                  Table~\ref{table:forwardAD}: because we compute adjoints, rather than starting with the root
                  variables, we start with the output and then work our way back to the roots (hence the term
                  ``reverse''). One reverse mode sweep gives us the full gradient.
                  The input and outputs of the trace are highlighted in gray.}}
     \label{table:reverseMode}
 \end{table}
 
%  \pagebreak
%  \section*{\sffamily \Large Mathematical Implementation} \label{Math}
%   
%   To write an effective implementation of AD, several mathematical considerations must made.

   \subsection*{\sffamily \large Forward or reverse mode?}

   For a function $f : \mathbb R^n \to \mathbb R^m$, suppose we wish to compute
   \textit{all} the elements of the $m \times n$ Jacobian matrix: which mode of AD should we use?
   Ignoring the overhead of building the expression graph,
   reverse mode, which requires $m$ sweeps, performs better when $n > m$.
   With the relatively small overhead, the performance of reverse-mode AD is superior when $n >> m$,
   that is when we have many inputs and few outputs. 
   This is the case when we map  many model parameters to an 
   objective function, and makes reverse mode highly applicable to high-dimensional modeling.

   However, if $n \le m$ forward mode performs better. Even when we have a comparable number of 
   outputs and inputs, as was the case in the log-normal density example, forward mode can be more 
   efficient, since there is less overhead associated with storing the expression graph in memory
   in forward than in reverse mode.
   \cite{BaydinEtAl:2015} compare the run time of forward and reverse-mode AD when applied to a statistical mechanics problem.
   The differentiated function is a many to one map, that is $f: \mathbb R^n \to \mathbb R$.
   For $n = 1$, forward-mode AD proves more efficient, but the result flips as $n$ increases (Table~\ref{tab:FwdRev}). %\\ \ \\
   
   \begin{table}
   \begin{center}
     \begin{tabular} {l c c c c}
     \rowcolor[gray]{0.95}  Dimension of input, $n$ & 1 & 8 & 29 & 50 \\ 
     Relative runtime & 1.13 & 0.81 & 0.45 & 0.26 \\
     \end{tabular}
     \caption{Relative runtime to compute a gradient with reverse-mode, when compared to forward-mode.
     \textit{The table summarizes results from an experiment conducted by \protect\cite{BaydinEtAl:2015}.
     The runtimes are measured by differentiating $f:\mathbb R^n \to \mathbb R$.
     Note forward-mode AD is more efficient when $n = 1$, but the result flips as we increase $n$.}}
     \label{tab:FwdRev}
   \end{center}
   \end{table}

   While a certain mode may overall work best for a function, we can also look at intermediate functions that get called inside a target function. 
   Such intermediate functions may be best differentiated using another mode of AD. 
   \cite{Phipps:2012} calculate derivatives of  intermediate expressions with reverse mode in an overall forward mode AD framework.
   They consider two forward-mode implementations; both use \textit{expression templates},
   a computational technique we will discuss in a few sections,
   and one of them uses \textit{caching}.
   Caching stores the values and partial derivatives of intermediate expressions,
   and insure they only get evaluated once.
   As a case study, \cite{Phipps:2012} consider an application to fluid mechanics,
   and report a 30\% reduction in runtime with intermediate reverse-mode, compared to standard forward AD without caching.
   The performance gain when forward AD already uses caching is however minor (less than 10\%).

   Another example where forward and reverse mode AD are both applied 
   is the computation of higher-order derivatives such as Hessians or Hessian vector products
   (see for example \cite{Pearlmutter:1994}). We will dwell a little longer on this in our discussion on higher-order derivatives.
   
  \section*{\sffamily \Large Computational implementation} \label{sec:computational}

  A computationally naive implementation of AD can result in prohibitively slow code and excess use of memory.
  Careful considerations can mitigate these effects.
  
  There exist several studies which measure the performance of different implementation schemes but these studies present limitations.
  First, they are indirect tests, as they often compare packages which have several distinguishing features.
  Secondly, computer experiments usually focus on a limited class of problems;
  there exists no standard set of problems spanning the diversity of AD applications
  and the development of a package is often motivated by a specific class of problems.
  Nevertheless, these test results work well as proof of concepts, and are aggregated in our study of computational techniques.

  Our focus is mainly on \texttt{C++} libraries, which have been compared to one another by \cite{Hogan:2014} and \cite{Carpenter:2015}.
  A common feature is the use of \textit{operator overloading} rather than \textit{source transformation}, two methods we will discuss in the upcoming sections.
  Beyond that, these libraries exploit different techniques, which will be more or less suited depending on the problem at hand.
  A summary of computational techniques is provided at the end of this section (Table~\ref{Tab:CompTechniques}).
  
  \subsection*{\sffamily \large Source transformation} \label{sec:sourceTransformation}

  A classic approach to compute and execute AD is \textit{source transformation},  
  as implemented for example in the \texttt{Fortran} package 
  \texttt{ADIFOR} \cite{Bischof:1996} and the \texttt{C++} packages \texttt{TAC++} \cite{Vossbeck:2008}
  and \texttt{Tapenade} \cite{Hascoet:2013}.
  We start with the source code of a computer program that implements our target function.
  A preprocessor then applies differentiation rules to the code, as prescribed by elementary operators and the chain rule, and
  generates new source code, which calculates derivatives. 
  The source code for the evaluation and the one for differentiation are then compiled and executed together.

  % The benefit of this method is that it explicitly spells out the code that evaluates the derivatives.
  % This makes the program very transparent and easy to optimize, either by hand or with a compiler.
  % \cite{Vossback:2008} run some performance tests on their package \texttt{TAC++},
  % though their paper does not benchmark performance against packages using operator overloading. 

  A severe limitation with source transformation is that it can only use information available at compile time,
  which prevents it from handling more sophisticated programming statements, such as while loops, \texttt{C++} templates,
  and other object-oriented features \cite{Vossbeck:2008, Hogan:2014}.
  There exist workarounds to make source transformation more applicable,
  but these end up being similar to what is done with operator overloading, albeit significantly less elegant \cite{Bischof:2000}.
  
  Moreover, it is commonly accepted in the AD community that source transformation works well for \texttt{Fortran} or \texttt{C}.
  \cite{Hascoet:2013} argue it is the choice approach for reverse-mode AD,
  although they also admit that, from a developer's perspective, implementing a package that deploys source transformation
  demands a considerable amount of effort.
  Most importantly, source transformation  cannot handle typical \texttt{C++} features. 
  For the latter, operator overloading is the appropriate technology. 

  % discuss \cite{Bischof:2000} % Bischol and Bucker 2000, COMPUTING DERIVATIVES OF COMPUTER PROGRAMS

  % \subsection*{\sffamily \large Operator overloading} \label{sec:operatorOverloading}
  \subsection*{\sffamily \large Operator overloading} \label{sec:operatorOverloading}

  Operator overloading is implemented in many packages, including the \texttt{C++}
  libraries:
  \texttt{Sacado} \cite{Gay:2005}, \texttt{Adept} \cite{Hogan:2014}, \texttt{Stan Math} \cite{Carpenter:2015},
  and \texttt{CoDiPack} \cite{Sagebaum:2017}.
  The key idea is to introduce a new class of objects, which contain the value of a variable on the expression
  graph and a \textit{differential component}.
  Not all variables on the expression graph will belong to this class.
  But the root variables, which require sensitivities, and all the intermediate variables which depend
  (directly or indirectly) on these root variables, will.
  In a forward mode framework, the differential component is the derivative with respect to one input. 
  In a reverse mode framework, it is the adjoint with respect to one 
  output\footnote{Provided we use, for forward and reverse mode, respectfully the right initial tangent and cotangent vector.}. 
  Operators and math functions are overloaded to handle these new types.
  We denote standard and differentiable types \texttt{real} and \texttt{var} respectively.

   \subsubsection*{\sffamily \large Memory management}

   AD requires  care when managing memory, particularly in our view when doing reverse-mode differentiation
   and implementing operator overloading.
 
   % Consider first a scalar function $f:\mathbb R^n \to \mathbb R$. 
   In forward mode, one sweep through the expression graph computes both the
   numerical value of the variables and their differential component with respect to an initial tangent.   
   The memory requirement for that sweep is then simply twice the requirement for a function evaluation.
   What is more, a \texttt{var} object, which corresponds to a node on the expression graph,
   needs only to be stored in memory temporally.
   Once all the nodes connected to that object have been evaluated, the \texttt{var} object can be discarded.

   When doing a reverse mode sweep, matters are more complicated
   because the forward evaluation trace happens first, and only then does the reverse AD trace occur.
   To perform the reverse trace, we need to access (i) the expression graph and (ii) the numerical values of the intermediate variables
   on that expression graph, required to calculate derivatives. These need to be stored in a persistent memory arena or \textit{tape}.
   Note further that neither (i) nor (ii) is known at compile time if the program we differentiate includes loops and conditional statements,
   and, in general, the memory requirement is dynamic.
  
   \subsubsection*{Retaping}
   
   In certain cases, a tape can be applied to multiple AD sweeps.
   An intuitive example is the computation of an $n \times m$ Jacobian matrix, where $n > 1$ and $m > 1$.
   Rather than reconstructing the expression graph and reevaluating the needed intermediate variables at each sweep,
   we can instead store the requisite information after the first sweep, and reuse it.
   Naturally, this strategy only works because the expression graph and the intermediate variables do not change from sweep to sweep.
   
   If we evaluate derivatives at multiple points, the intermediate variables almost certainly need to be reevaluated and, potentially, stored in memory.
   On the other hand, the expression graph may not change.
   Thence a single tape for the expression graph can be employed to compute AD sweeps at multiple points. 
   This scheme is termed \textit{retaping}; implementations can be found in \texttt{ADOL-C} \cite{Griewank:1999, Walther:2012}
   and \texttt{CppAD} \cite{Bell:2012}.
 
   The conservation of the expression graph from point to point does however not hold in general.
   Our target function may contain conditional statements, which control the expression graph
   our program generates. In such cases, new conditions require new tapes.
   Retaping can then be used, provided we update the expression graph as soon as the control flow changes.
%   One could also conceive storing multiple graphs, corresponding to multiple control flows,
%   but it is unclear how well such a scheme would scale for softwares that support the differentiation
%   of arbitrary functions specified by a user.

   \subsubsection*{Checkpoints}

   \textit{Checkpointing} reduces peak memory usage and more generally trades runtime and memory cost.
   Recall our target function $f$ is a composite of, say, $L$ simpler functions $f^l$.
   Combining equations~\ref{eq:chainJacobian} and \ref{eq:reverse},
   the action of the transpose Jacobian matrix, $w'$, computed by one reverse sweep of AD verifies
   \begin{eqnarray}
     \begin{aligned}
            w' & = J^T \cdot w \\
                & = (J_L \cdot J_{L - 1} \cdot ... \cdot J_1 \cdot)^T \cdot w \\
                % & = J_1^T \cdot J_2^T \cdot ... \cdot J_L^T \cdot \bar u
     \end{aligned}
   \end{eqnarray}
   where $w$ is the initial cotangent. We can split $w'$ into a sequence of $K$ components
   \begin{eqnarray}
     \begin{aligned}
     w' & = (J_{\phi(1)} \cdot ... \cdot J_1)^T w_1 \\
     w_1 & = (J_{\phi(2)} \cdot ... \cdot J_{\phi(1) + 1})^T w_2 \\
     . . . \\
     w_{K - 1} &= (J_{\phi(K - 1)} \cdot ... \cdot J_{\phi(K - 2) + 1})^T w
     \end{aligned}
     \label{eq:Kdirections}
   \end{eqnarray}
   which can be sequentially evaluated by a reverse AD sweep, starting from the last line.
   The index function $\phi$ tells us where in the composite function the splits occur.
   
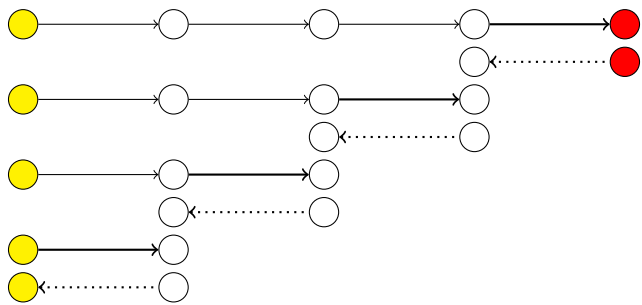
\begin{figure}[htbp]
  \begin{center}
  \begin{tikzpicture}
  [
    Checkpoint/.style={circle, draw=black!, fill=white!35, thin, minimum size=0.5mm},
    Input/.style={circle, draw=black!, fill=yellow!255, thin, minimum size=0.5mm},
    Output/.style={circle, draw=black!, fill=red!255, thin, minimum size=0.5mm}
  ]
  % Nodes
  % \node[Gray] (AD) at(0, 0) {\ \ \large \textbf{AD} \ \ };
  
 % first layer
 \node[Input] (in) at(0, 0){};
 \node[Checkpoint] (c1) at(2, 0){};
 \node[Checkpoint] (c2) at(4, 0){};
 \node[Checkpoint] (c3) at(6, 0){};
 \node[Output] (out) at (8, 0){};
 
 % second layer
  \node[Checkpoint] (c3-2) at(6, -0.5){};
 \node[Output] (out-2) at (8, -0.5){};
 
 % third later
  \node[Input] (in-3) at(0, -1){};
 \node[Checkpoint] (c1-3) at(2, -1){};
 \node[Checkpoint] (c2-3) at(4, -1){};
 \node[Checkpoint] (c3-3) at(6, -1){};
 
  % fourth layer
  \node[Checkpoint] (c2-4) at(4, -1.5){};
  \node[Checkpoint] (c3-4) at (6, -1.5){};
  
  % fifth later
  \node[Input] (in-5) at(0, -2){};
  \node[Checkpoint] (c1-5) at(2, -2){};
  \node[Checkpoint] (c2-5) at(4, -2){};
  
  % sixth layer
  \node[Checkpoint] (c1-6) at(2, -2.5){};
  \node[Checkpoint] (c2-6) at(4, -2.5){};
  
  % seventh layer
  \node[Input] (in-7) at(0, -3){};
  \node[Checkpoint] (c1-7) at(2, -3){};  
 
 %eighth layer
 \node[Input] (in-8) at(0, -3.5){};
 \node[Checkpoint] (c1-8) at(2, -3.5){};  

%  \node[Box, text width = 6cm] (goal) at(10, 0) {\noindent \begin{center} Differentiate math \\ functions and algorithms \end{center} \ \ };
                  
  % Lines for first layer
  \path [->, draw, thin] (in) -- (c1);
  \path [->, draw, thin] (c1) -- (c2);
  \path [->, draw, thin] (c2) -- (c3);
  \path [->, draw, thick] (c3) -- (out);
  
  % Lines for second layer
  \path [<-, draw, thick, dotted] (c3-2) -- (out-2);
  
  % Third layer
  \path [->, draw, thin] (in-3) -- (c1-3);
  \path [->, draw, thin] (c1-3) -- (c2-3);
  \path [->, draw, thick] (c2-3) -- (c3-3);
  
  % Fourth layer
  \path [<-, draw, thick, dotted] (c2-4) -- (c3-4);
  
   % fifth layer
  \path [->, draw, thin] (in-5) -- (c1-5);
  \path [->, draw, thick] (c1-5) -- (c2-5);
  
  % sixth layer
  \path [<-, draw, thick, dotted] (c1-6) -- (c2-6);
  
  % seventh layer
  \path [->, draw, thick] (in-7) -- (c1-7);
  
  %eighth layer
  \path [<-, draw, thick, dotted] (in-8) -- (c1-8);
  
  \end{tikzpicture}
  \end{center}
  \caption{Checkpointing with the recompute-all approach. \textit{In the above sketch, the target function
  is broken into 4 segments, using three checkpoints (white nodes). The yellow and red nodes
  respectively represent the input and output of the function.
  During the forward sweep, we only record when the arrow is thick.
  The dotted arrow represents a reverse AD sweep.
  After each reverse sweep, we start a new forward evaluation sweep from the input.}}
  \label{fig:checkpoint1}
  \end{figure}

   Correspondingly, the program that implements $f$ can be split into $K$ sections,
   with a \textit{checkpoint} between each section.
   We then have several options, the first one being the \textit{recompute-all} approach.
   Under this scheme, the program begins with a forward evaluation sweep, but here is the catch:
   we do not record the expression graph and the intermediate values, 
   as we would in previously discussed implementations of reverse-mode AD,
   until we reach the last checkpoint.
   That is we only record from the last checkpoint onward.
   When we begin the reverse sweep, the size of the tape is relatively small,
   thence the reduction in peak memory cost.
   Of course, we can only perform the reverse sweep from the output to the last checkpoint.
   Once we reach the latter, we rerun the forward evaluation sweep from the beginning, and only record
   between the second-to-last and the last checkpoint, which gives us the requisite information for the next reverse sweep. 
   And so on.
   The reason we call this approach \textit{recompute-all} is because we rerun a partial forward evaluation trace,
   with partial recording, for each checkpoint (figure~\ref{fig:checkpoint1}).
   Moreover, the reduction in peak memory comes at the cost of doing multiple forward evaluation traces.
   
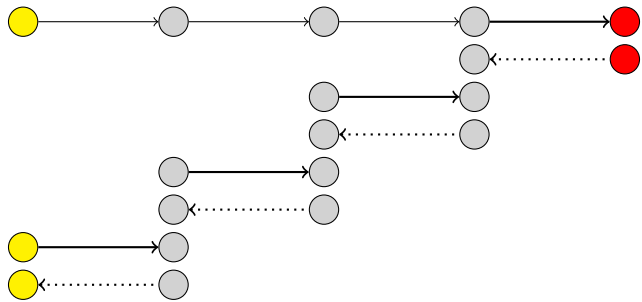
\begin{figure}[htbp]
  \begin{center}
  \begin{tikzpicture}
  [
    Checkpoint/.style={circle, draw=black!, fill=gray!35, thin, minimum size=0.5mm},
    Input/.style={circle, draw=black!, fill=yellow!255, thin, minimum size=0.5mm},
    Output/.style={circle, draw=black!, fill=red!255, thin, minimum size=0.5mm}
  ]
  % Nodes
  % \node[Gray] (AD) at(0, 0) {\ \ \large \textbf{AD} \ \ };
  
 % first layer
 \node[Input] (in) at(0, 0){};
 \node[Checkpoint] (c1) at(2, 0){};
 \node[Checkpoint] (c2) at(4, 0){};
 \node[Checkpoint] (c3) at(6, 0){};
 \node[Output] (out) at (8, 0){};
 
 % second layer
  \node[Checkpoint] (c3-2) at(6, -0.5){};
 \node[Output] (out-2) at (8, -0.5){};
 
 % third later
 \node[Checkpoint] (c2-3) at(4, -1){};
 \node[Checkpoint] (c3-3) at(6, -1){};
 
  % fourth layer
  \node[Checkpoint] (c2-4) at(4, -1.5){};
  \node[Checkpoint] (c3-4) at (6, -1.5){};
  
  % fifth later
  \node[Checkpoint] (c1-5) at(2, -2){};
  \node[Checkpoint] (c2-5) at(4, -2){};
  
  % sixth layer
  \node[Checkpoint] (c1-6) at(2, -2.5){};
  \node[Checkpoint] (c2-6) at(4, -2.5){};
  
  % seventh layer
  \node[Input] (in-7) at(0, -3){};
  \node[Checkpoint] (c1-7) at(2, -3){};  
 
 %eighth layer
 \node[Input] (in-8) at(0, -3.5){};
 \node[Checkpoint] (c1-8) at(2, -3.5){};  

%  \node[Box, text width = 6cm] (goal) at(10, 0) {\noindent \begin{center} Differentiate math \\ functions and algorithms \end{center} \ \ };
                  
  % Lines for first layer
  \path [->, draw, thin] (in) -- (c1);
  \path [->, draw, thin] (c1) -- (c2);
  \path [->, draw, thin] (c2) -- (c3);
  \path [->, draw, thick] (c3) -- (out);
  
  % Lines for second layer
  \path [<-, draw, thick, dotted] (c3-2) -- (out-2);
  
  % Third layer
  \path [->, draw, thick] (c2-3) -- (c3-3);
  
  % Fourth layer
  \path [<-, draw, thick, dotted] (c2-4) -- (c3-4);
  
   % fifth layer
  \path [->, draw, thick] (c1-5) -- (c2-5);
  
  % sixth layer
  \path [<-, draw, thick, dotted] (c1-6) -- (c2-6);
  
  % seventh layer
  \path [->, draw, thick] (in-7) -- (c1-7);
  
  %eighth layer
  \path [<-, draw, thick, dotted] (in-8) -- (c1-8);

  \end{tikzpicture}
  \end{center}
  \caption{Checkpointing with the store-all approach. \textit{
  This time, the forward sweep records intermediate values at each checkpoint.
  The forward sweep can then be rerun between two checkpoints, as opposed to starting from the input.}}
  \label{fig:checkpoint2}
  \end{figure}

   It may be inefficient to rerun forward sweeps from the start every time. 
   We can instead store the values of intermediate variables at every checkpoint,
   and then perform forward sweeps only between two checkpoints (figure~\ref{fig:checkpoint2}).
   Under this scheme, the forward evaluation trace is effectively run twice. 
   This variant of checkpointing is termed \textit{store-all}.
   Storing intermediate results at each checkpoint may for certain problems be too memory expensive.
   Fortunately, recompute-all and store-all constitute two ends of a spectrum: in between methods can be used
   by selecting a subset of the checkpoints where the values are stored.
   Such checkpoints constitute \textit{snapshots}, and give us more control on peak memory usage.
   
   A natural question is where should we place the checkpoints?
   Ideally, we would like, given a memory constraint, to reduce runtime.
   There is in general, as argued by \cite{Hascoet:2013}, no optimal placement and users should be given the
   freedom to place checkpoints by hand.   
   In the case where we can split the the program for $f$ into sequential \textit{computational steps},
   with comparable computational costs, placing the checkpoints according to a
   \textit{binomial partitioning} scheme \cite{Griewank:1992} produces an optimal placement \cite{Grimm:1996}.
   
   There are several packages which implement checkpointing: 
   for example \texttt{Adol-C} and \texttt{CppAD} couple the method with operator overloading,
   while it is used with source transformation in \texttt{Tapenade}.
   
   We conclude this section by drawing a connection between checkpointing and parallelizing AD.
   Suppose a section of $f$ can be split into segments we can evaluate in parallel.
   Specifically, each segment is evaluated on a separate computer core.
   Conceptually, we can place checkpoints around these segments, and create separate tapes for each of these segments.
   As a result, memory is kept local and does not need to be shared between cores.
   An example of such a procedure is the \textit{map reduce} functionality in \texttt{Stan Math},
   describe in the \textit{Stan book} \cite{StanBook}. Note this routine is not explicitly described as a checkpointing method.

  \subsubsection*{Region Based Memory}
  
   If we allocate and free memory one intermediate variable at a time, the overhead cost becomes quite severe.
   A better alternative is to do region-based memory management (see for example \cite{Gay:2001}).
   Memory for the calculation of derivatives is allocated element wise on a custom stack.
   Once the derivatives have been evaluated, the entire stack is destroyed and the associated memory freed.
   This spares the program the overhead of freeing memory one element at a time.
   \cite{Gay:2005} implements this technique in the \texttt{C++} package \texttt{RAD} (\textit{Reverse AD}, the reverse mode version of \texttt{Sacado}),
   and benchmarks its performance against the \texttt{C}/\texttt{C++} package \texttt{ADOL-C}.
   The results show \texttt{RAD} runs up to twice as fast, when differentiating quality mesh functions with varying levels of complexity.
 
   \subsection*{\sffamily Expression templates}
   % \subsection*{\sffamily  Expression templates}

   Rather than overloading an operator such that it returns a custom object,
   we can code the operator to return an \textit{expression template} \cite{Veldhuizen:1995}. 
   This technique, which builds on lazy evaluation, is widely used in packages for vector and matrix operations and was first 
   applied to AD in the \texttt{Fad} package \cite{Aubert:2001}. 
   \cite{Phipps:2012} recommend several techniques to improve expression templates and showcase its implementation in \texttt{Sacado}.
   \cite{Hogan:2014} and more recently \cite{Sagebaum:2017} apply the method to do full reverse-mode AD,
   respectively in \texttt{Adept} and \texttt{CoDiPack}.
   
   Under the expression template scheme, operators return a specialized object that characterizes an intermediate variable as a node in the expression graph.
   As an example, consider the following programing statement:
   %
%   \begin{algorithm}[H]
%     \caption*{ \textit{Code to compute $f(x, a) = a \cos(x)$, when we require sensitivity for $x$. }}
     \\ \ \\
     \texttt{
     var x; \\
     real a;  \\
     var b = cos(x);  \\
     var c = b * a; \\ } \\
  % \end{algorithm}
   %
   \noindent
   Take the \textit{multiply} or \texttt{*} operator, which is a map $f:\mathbb R^2 \to \mathbb R$.
   Rather than returning a \texttt{var} object, we overload \texttt{*} to return a template object of type
   \mbox{\texttt{multiply<T\_a, T\_b>}}, where \texttt{T\_a} and \texttt{T\_b} designate the types of
   $a$ and $b$. \texttt{T\_a} or \texttt{T\_b} can be \texttt{real}, \texttt{var}, or expression
   templates. In the above example, $b =  \cos x$, $x$ is a \texttt{var}, and $a$ a \texttt{real}.
   Then $c$ is assigned an object of type \mbox{\texttt{multiply<cos<var>, real>}}, rather than
   an evaluated \texttt{var} object (Table~\ref{tab:templateExpression}).
   The nodes in the expression graph  hence contain references. An optimizing compiler can
   then combine the nodes to generate more efficient code, which would be functionally equivalent
   to: \\ \ \\
   \texttt{
     c.val\_ = cos(x).val\_ * a; \\
     x.adj\_ = c.adj\_ * a * (- sin(x).val\_) \\
     } \\
   where \texttt{val\_} and \texttt{adj\_} respectively denote the value and the adjoint stored in 
   the \texttt{var} object. This operation is called \textit{partial evaluation}. Notice that \texttt{b} does not appear
   in the above programing statement. Moreover, the use of expression templates removes temporary 
   \texttt{var} objects, which saves memory and eliminates the overhead associated with constructing 
   intermediate variables. Furthermore, the code can eliminate redundancies, such as the addition and
   later subtraction of one variable.

   \begin{table}[H]
     \centering
     \begin{tabular}{c l}
     \textbf{variable} & \textbf{type} \\
     %\hline
     \texttt{x} & \texttt{var} \\
     \texttt{a} & \texttt{real} \\
     \texttt{b} & \texttt{cos<var>} \\
     \texttt{c} & \texttt{multiply<cos<var>, real>}\\
     %\hline \hline
     \rowcolor[gray]{0.95} \textbf{Algorithm} & \texttt{b = cos(x);} \\
     \rowcolor[gray]{0.95}                & \texttt{c = b * a;} \\
     \end{tabular}
     \caption{Variable types when using expression templates}
     \label{tab:templateExpression}
   \end{table}

   \cite{Hogan:2014} evaluates the benefits of coupling expression templates and AD by benchmarking \texttt{Adept} 
   against other AD packages that use operator overloading, namely \texttt{Adol-C}, \texttt{CppAD}, and \texttt{Sacado}.
   His computer experiment focuses on an application in fluid dynamics, using a simple linear scheme and a more complex nonlinear method.
   In these specific instances, \texttt{Adept} outperformances other packages, both for forward and reverse mode AD: 
   it is 2.6 - 9 times faster and 1.3 - 7.7 times more memory efficient, depending on the problem and the benchmarking package at hand.
   \cite{Carpenter:2015} corroborate the advantage of using expression templates by testing two implementations of a function that
   returns the log sum of exponents.
   The first implementation looks as follows: \\ \ \\
   \texttt{
     for (int = 0; i < x.size(); i++) \\
     \phantom{xx} total = log(exp(total) + exp(x(i))); \\
     return total;
   } \\ \ \\
  This code is inefficient because the log function gets called multiple times, when it could be called once, as below: \\ \ \\
  \texttt{
    for (int  = 0; i < x.size(); i++) \\
    \phantom{xx} total += exp(x(i)); \\
    return log(total);
  } \\ \ \\
  \texttt{Stan Math} requires 40\% less time to evaluate and differentiate the function, when we use the second implementation instead of the first one.
  On the other hand \texttt{Adept} achieves optimal speed with both implementations.
  When applied to the second algorithm, both packages perform similarly.
  Hence, while careful writing of a function can match the results seen with expression templates, this requires more coding effort from the user.  

  \begin{table}
   \footnotesize{
   \renewcommand{\arraystretch}{0.8}
   \begin{tabular}{l l l l}
   \textbf{Method} & \textbf{Benefits} & \textbf{Limitations} & \textbf{Packages} \\
   % \hline
   \rowcolor[gray]{0.95} \textbf{Source}             & Can be optimized by hand or with & Can only differentiate functions          & \texttt{AdiFor}, \texttt{TAC++} \\
   \rowcolor[gray]{0.95} \textbf{transformation} &  a compiler.                         & which are fully defined at compile &    \texttt{Tapenade} \\
   \rowcolor[gray]{0.95}          & & time. &  \vspace{2mm} \\
   \textbf{Operator}         & Can handle most programing & Memory handling requires & All the packages\\
   \textbf{Overloading}    & statements. & some care. & listed below. \vspace{2mm} \\
   Retaping                     & Fast when we differentiate through    & Applicable only if the expression    & \texttt{Adol-C}, \texttt{CppAd} \\
                                      & the same expression graph& graph is conserved from point\\
                                      & multiple times.              & to point.  When the graph \\
                                      &                                     &  changes a new tape can be \\
                                      &                                     & created. \vspace{1.5mm} \\
   Checkpointing & Reduces peak memory usage. & Longer runtime. An optimal   & \texttt{Adol-C}, \texttt{CppAD} \\
                          &                                                   &  placement is binary partition, but  & (\texttt{Tapenade} for source \\
                          &                                                   &  this scheme does not always    &  transformation). \\
                          &                                                   & apply.     \\
   Region-based  & Improves speed and memory. & Makes the implementation less  & \texttt{Adept}, \texttt{Sacado}, \\
   memory      &  &  transparent. Some care is & \texttt{Stan Math}  \\
     & & required for nested AD. & \vspace{2mm} \\
   Expression  & Removes certain redundant operations,  &   & \texttt{Adept}, \texttt{CoDiPack} \\ 
   templates    & whereby the code becomes faster and  & & \texttt{Sacado} (with \\
                       & more memory efficient, and allows & & some work). \\ 
                       & user to focus on code clarity. \vspace{2mm} \\
   \hline
  \end{tabular}
  }  % footnotesize
    \caption{Summary of computation techniques to optimize AD.
    \textit{The right most column lists packages discussed in this section that implement a given computational method.
              This is by no means a comprehensive survey of all AD packages.
              A further word of caution: when choosing a package, users should also look at other criterions,
              such as ease of use, extensibility, and built-in mathematical functions.}}
    \label{Tab:CompTechniques}
  \end{table}

%  \iffalse  %%% MARKER WHERE EVERYTHING WORKS!!!!!

  \section*{Mathematical implementation}
  
  We now study several mathematical considerations users can make to optimize the differentiation of a complex function.
  The here discussed techniques are generally agnostic to which AD package is used;
  they may however already be implemented as built-in features in certain libraries.
  We exemplify their use and test their performance using \texttt{Stan Math}
  and reverse-mode AD\footnote{Similar results are expected with forward mode.}.
  The driving idea in this section is to gain efficiency by reducing the size of the expression graph generated by the code.
  
  \subsection*{Reducing the number of operations in a function}
  
  A straightforward way of optimizing code is to minimize the number of operations in a function.
  This can be done by storing the result of calculations that occur multiple times inside intermediate variables.
  When evaluating a function, introducing intermediate variables reduces runtime but increases memory usage.
  With AD, eliminating redundant operations also reduces the generated expression graph,
  which means intermediate variables both improve runtime and memory usage.
 
  Consider for example the implementation of a function that returns a $2 \times 2$ matrix exponential.
  The matrix exponential extends the concept of a scalar exponential to matrices 
  and can be used to solve linear systems of ordinary differential equations (ODEs).
  There exist several methods to approximate matrix exponentials, the most popular one being the Pad\'e approximation, coupled with scaling and squaring; 
  for a review, see \cite{Moler:2003}.
  For the $2 \times 2$ matrix, 
  \begin{eqnarray*}
    A = \left[\begin{array}{cc}
	a & b \\
	c & d
	\end{array}\right]
  \end{eqnarray*}
  the matrix exponential can be worked out analytically (see \cite{ToddWeisstein}). 
  Requiring \mbox{$\Delta := \sqrt{(a - d)^2 + 4bc} $} to be real, we have:
  \begin{eqnarray}
    \exp(A) & = & \frac{1}{\Delta} \left[\begin{array}{cc} m_{11} & m_{12} \\ m_{21} & m_{22} \end{array} \right]
  \label{Eq:2x2MatrixExp}
  \end{eqnarray}
  where
  \begin{eqnarray*}
    m_{11} & =  & e^{(a + d) / 2} [ \Delta \cosh(\Delta / 2) + (a - d) \sinh(\Delta / 2) ] \\
    m_{12} & = & 2 b e^{(a + d) / 2} \sinh(\Delta / 2) \\
    m_{21} & = & 2 c e^{(a + d) / 2} \sinh(\Delta / 2) \\
    m_{22} & = & e^{(a + d) / 2} [\Delta \cosh(\Delta / 2) + (d - a) \sinh(\Delta / 2)]
  \end{eqnarray*}
  The above expression only contains elementary operators, and can be differentiated by any standard AD library.
  A  ``direct" implementation in \texttt{C++} might simply translate the above mathematical equations
  and look as follows:
  \begin{algorithm}[H]
  \caption*{\textit{Standard implementation of $2 \times 2$ matrix exponential}}
    \footnotesize{\texttt{
    \ \\
      T a = A(0, 0), b = A(0, 1), c = A(1, 0), d = A(1, 1), delta; \vspace{2mm} \\ 
      delta = sqrt(square(a - d) + 4 * b * c); \vspace{2mm} \\
     Matrix<T, Dynamic, Dynamic> B(2, 2); \\
     B(0, 0) = exp(0.5 * (a + d)) * (delta * cosh(0.5 * delta) + (a - d) * sinh(0.5 * delta)); \\
     B(0, 1) = 2 * b * exp(0.5 * (a + d)) * sinh(0.5 * delta); \\
     B(1, 0) = 2 * c * exp(0.5 * (a + d)) * sinh(0.5 * delta);  \\
    B(1, 1) = exp(0.5 * (a + d)) * (delta * cosh(0.5 * delta) + (d - a) * sinh(0.5 * delta)); \vspace{2mm} \\
    return B / delta; \\
    }}
  \end{algorithm}
  \noindent
  Note we use the matrix library \texttt{Eigen} \cite{Eigen:2013}. 
  \texttt{T} represents a template type which, in need, will either be a \texttt{real} or a \texttt{var}.
  An optimized implementation removes redundant operations and introduces variables to store intermediate results:
    \begin{algorithm}[H]
  \caption*{\textit{Optimized implementation of $2 \times 2$ matrix exponential}}
    \footnotesize{\texttt{
    \ \\
  T a = A(0, 0), b = A(0, 1), c = A(1, 0), d = A(1, 1), delta; \\
  delta = sqrt(square(a - d) + 4 * b * c);  \vspace{2mm} \\
  Matrix<T,  Dynamic, Dynamic> B(2, 2); \\
  T half\_delta = 0.5 * delta; \\
  T cosh\_half\_delta = cosh(half\_delta); \\
  T sinh\_half\_delta = sinh(half\_delta); \\
  T exp\_half\_a\_plus\_d = exp(0.5 * (a + d));  \\
  T Two\_exp\_sinh = 2 * exp\_half\_a\_plus\_d * sinh\_half\_delta;  \\
  T delta\_cosh = delta * cosh\_half\_delta;  \\
  T ad\_sinh\_half\_delta = (a - d) * sinh\_half\_delta;  \vspace{2mm}  \vspace{2mm} \\
  B(0, 0) = exp\_half\_a\_plus\_d * (delta\_cosh + ad\_sinh\_half\_delta); \\
  B(0, 1) = b * Two\_exp\_sinh;  \\
  B(1, 0) = c * Two\_exp\_sinh;  \\
  B(1, 1) = exp\_half\_a\_plus\_d * (delta\_cosh - ad\_sinh\_half\_delta); \vspace{2mm} \\
  return B / delta; \\
  }}
  \end{algorithm}
  \noindent
  The two codes output the same mathematical expression, but the second implementation is about a third faster,
  as a result of producing an expression graph with less nodes (Table~\ref{tab:matrixExp}).
  This is measured by applying both implementations to 1000 randomly generated $2 \times 2$ matrices, labelled ``$A$".
  We compute the run time by looking at the wall time required to evaluate $\exp(A)$ and differentiate
  it with respect to every elements in $A$ (thereby producing 16 partial derivatives).
  The experiment is repeated 100 times to quantify the variation in the performance results.

  Unfortunately the optimized approach requires more coding effort, produces more lines of code, and ultimately hurts clarity,
  as exemplified by the introduction of a variable called \texttt{ad\_sinh\_half\_delta}.
  Whether the trade-off is warranted or not depends on the user's preferences. 
  Note expression templates do not perform the here-discussed optimization,
  as a compiler may fail to identify expressions which get computed multiple times inside a function.
  
  \begin{table}[H]
    \begin{center}
    \begin{tabular}{c c c c}
    \rowcolor[gray]{0.95} \textbf{Implementation} & \textbf{Number of nodes} & \textbf{Run time (ms)} &  \textbf{Relative run time} \\
    % \hline
    Standard & 41 & $2.107 \pm 0.35$ & 1.0 \\
    Optimized & 25 & $1.42 \pm 0.28$ & 0.68 \\
    \end{tabular}
    \end{center}
    \caption{Run time for a standard and an optimized implementation of a $2 \times 2$ matrix exponential.
    \textit{The optimized implementation of the matrix exponential removes redundant operations by 
    introducing new intermediate variables.
    As a result, the code produces an expression with less nodes, which leads to more efficient differentiation.}
    }
    \label{tab:matrixExp}
  \end{table}

  \subsection*{Optimizing with super nodes}

  There exist more sophisticated approaches to reduce the size of an expression graph.
  This is particularly useful, and for practical purposes even necessary, when differentiating iterative algorithms.
  Our target function, $f$, may embed an implicit function, $f^k$, such as the solution to an equation we solve numerically.
  Numerical solvers often use iterative algorithms to find such a solution.
  The steps of these algorithms involve simple operations, hence it is possible to split $f^k$ into elementary functions,
  here indexed 1, ..., $M$:
  \begin{eqnarray*}
    \begin{aligned}
    f & = f^L \circ  ... \circ f^{k+1} \circ f^k \circ ... \circ f^1 \\
      & = f^L \circ ... \circ f^{k + 1} \circ f^k_M \circ f^k_{M - 1} \circ ... \circ f^k_1 \circ ... \circ f^1
    \end{aligned}
  \end{eqnarray*}
  Note $M$, the number of elementary functions we split $f^k$ into, depends on the number of steps the numerical solver 
  takes, and hence varies from point to point.
  Using the above we can then easily deploy our usual arsenal of tools to do forward or reverse-mode AD.
  But doing so produces large expression graphs, which can lead to floating point precision errors,
  excessive memory usage, and slow computation.
  To mitigate these problems, we can elect to not split $f^k$, and use a custom method to compute $f^k$'s contribution
  to the action of the Jacobian matrix. 
  A more algorithmic way to think about this is to say we do not treat $f^k$ as a function generating a graph,
  but as a stand-alone operator which generates a node or \textit{super node}.
  A super node uses a specialized method to compute Jacobian matrices and often produces a relatively small expression graph.

  As a case study, we present several implementations of a numerical algebraic solver,
  and run performance tests to measure run time. Ease of implementation is also discussed.
  In particular, we examine a well-known strategy which consists in using the implicit function theorem.
  For a more formal discussion of the subject, we recommend \cite{Bell:2008} and chapter 15 of \cite{Griewank:2008}.
  % which also present methods to compute second-order derivatives.

  Another example where super nodes yield a large gain in performance is the numerical solving of ODEs,
  where the iterative algorithm is used to simultaneously evaluate a solution and compute sensitivities. 
  Section 13 of \cite{Carpenter:2015} provides a nice discussion on this approach.
  Finally, \cite{Giles:2008} summarizes several results to efficiently compute matrix derivatives,
  including matrix inverses, determinants, matrix exponentials, and Cholesky decompositions.

  \subsubsection*{Differentiating a numerical algebraic solver}

  We can use a numerical algebraic solver to find the root of a function and solve nonlinear 
  algebraic equations. Formally, we wish to find $y^* \in \mathbb{R}^N$ such that:
  \begin{eqnarray*}
    f(y^*, \theta) = 0
  \end{eqnarray*}
  where $\theta \in \mathbb{R}^K$  is a fixed vector of auxiliary parameters, or more generally a fixed vector of 
  variables for which we require sensitivities. That is we wish to compute the $N \times K$ Jacobian
  matrix, $J$, with $(i, j)^\mathrm{th}$ entry:
   \begin{eqnarray*}
     J_{i,j} = \frac{\partial y_i^*}{\partial \theta_j}(y^*, \theta)
   \end{eqnarray*}
  %
  %Regular AD is readily applicable to most numerical solvers and can be used to compute $J$.
  Consider, for demonstrating purposes, an implementation of Newton's method.
  For convenience, we note $J^y$ the Jacobian matrix, which contains the partial derivatives $f$
  with respect to $y$.

  \begin{algorithm}[H]
  \caption*{\textit{Standard iterative algorithm: Newton solver with a fixed step size}}
    \begin{enumerate}
      \item Start with an initial guess $y_0$.
      \item While a \textit{convergence criterion} has not been met:
      \begin{eqnarray*}
        y^{(i + 1)} = y^{(i)} -  [J^y(y^{(i)}, \theta)]^{-1} f(y^{(i)}, \theta)
      \end{eqnarray*}
      \item Return $y^{(I)}$, where $I$ is the number of steps required to reach convergence,
      or a warning message stating convergence could not be reached.
   \end{enumerate}
  \end{algorithm}
  \noindent
  %  The solver can have several tuning parameters. 
  %  In the described implementation, we simply use a tolerance level and a maximum number of steps.
  In the case where an analytical expression for $J^y$ is provided, we can readily apply regular AD to differentiate the algorithm.
  If we are working with overloaded operators, this simply means templating the solver so that it can handle \texttt{var} types.
  Unfortunately, if the number of steps required to reach convergence is high, the algorithm produces a large expression graph.
  The problem is exacerbated when $J^y$ is not computed analytically.
  Then AD must calculate the higher-order derivative:
  \begin{eqnarray*}
    \frac{\partial^2 f}{\partial y \partial \theta}
  \end{eqnarray*}
  Coding a method that returns such a derivative requires the implementation of \textit{nested} AD, 
  and a careful combination of the reverse and forward mode chain rule.
  % Most packages provide a routine that does this for Hessian matrices, though not for more ``arbitrary" higher-order differential operators.
  Tackling this problem with brute-force AD hence demands a great coding effort and is likely to be prohibitively slow.

  An alternative approach is to use the implicit function theorem which tells us that, under some regularity conditions
  in the neighborhood of the solution:
  \begin{eqnarray}
    J = - [J^y(y^*, \theta)]^{-1} J^\theta(y^*, \theta)
    \label{eq:jacobian}
  \end{eqnarray}
  where, consistent with our previous notation, $J^\theta$ is the Jacobian matrix of $f$ with respect to $\theta$.
  We have further made the dependence on $y$ and $\theta$ explicit. 
  Note the right-hand side is evaluated at the solution $y^*$.
 
  This does not immediately solve the problem at hand but it significantly simplifies it. 
  Computing $J^y(y^*, \theta)$ and $J^\theta(y^*, \theta)$ is, in practice, straightforward.
  In simple cases, these partials can be worked out analytically, but more generally, we can use AD.
  Many algorithms, such as Newton's method and Powell's dogleg solver \cite{Powell:1970} already compute the matrix $J^y$
  to find the root, a result that can then be reused.
  Note furthermore that with Equation~\ref{eq:jacobian}, calculating $J$ reduces to a first-order differentiation problem.
  
  Ideally, super nodes are already implemented in the package we plan to use.
  If this is not the case, the user can attempt to create a new function with a custom differentiation method.
  The amount of effort required to do so depends on the AD library at hand and the user's coding expertise;
  the paper further discusses this point in its section on \textit{Extensibility}.
  For now, our advice is to build new operators when brute force AD is prohibitively slow or when developing general purpose tools.

  To better inform the use of super nodes, we compare the performance of two dogleg solvers.
  The first one uses a templated version of the algorithm and applies regular AD.
  The second one uses Equation~\ref{eq:jacobian}.
  For simplicity, we provide $J^y$ in analytical form; for the second solver, $J^\theta$ is computed with AD.
  The solvers are applied to an archetypical problem in pharmacometrics, namely the modeling of steady states for patients undergoing a medical treatment.
  The details of the motivating problem are described in the appendix.
  The number of states in the algebraic equation varies from 2 to 28, and the number of parameters which require sensitivities respectively from 4 to 30.
  We use \texttt{Stan Math}'s built-in algebraic solver as a benchmark, a dogleg algorithm which also uses Equation~\ref{eq:jacobian}
  but computes $J^y$ with AD.
  
\begin{figure}
  \begin{center}
    \includegraphics[width=7cm, height = 7cm]{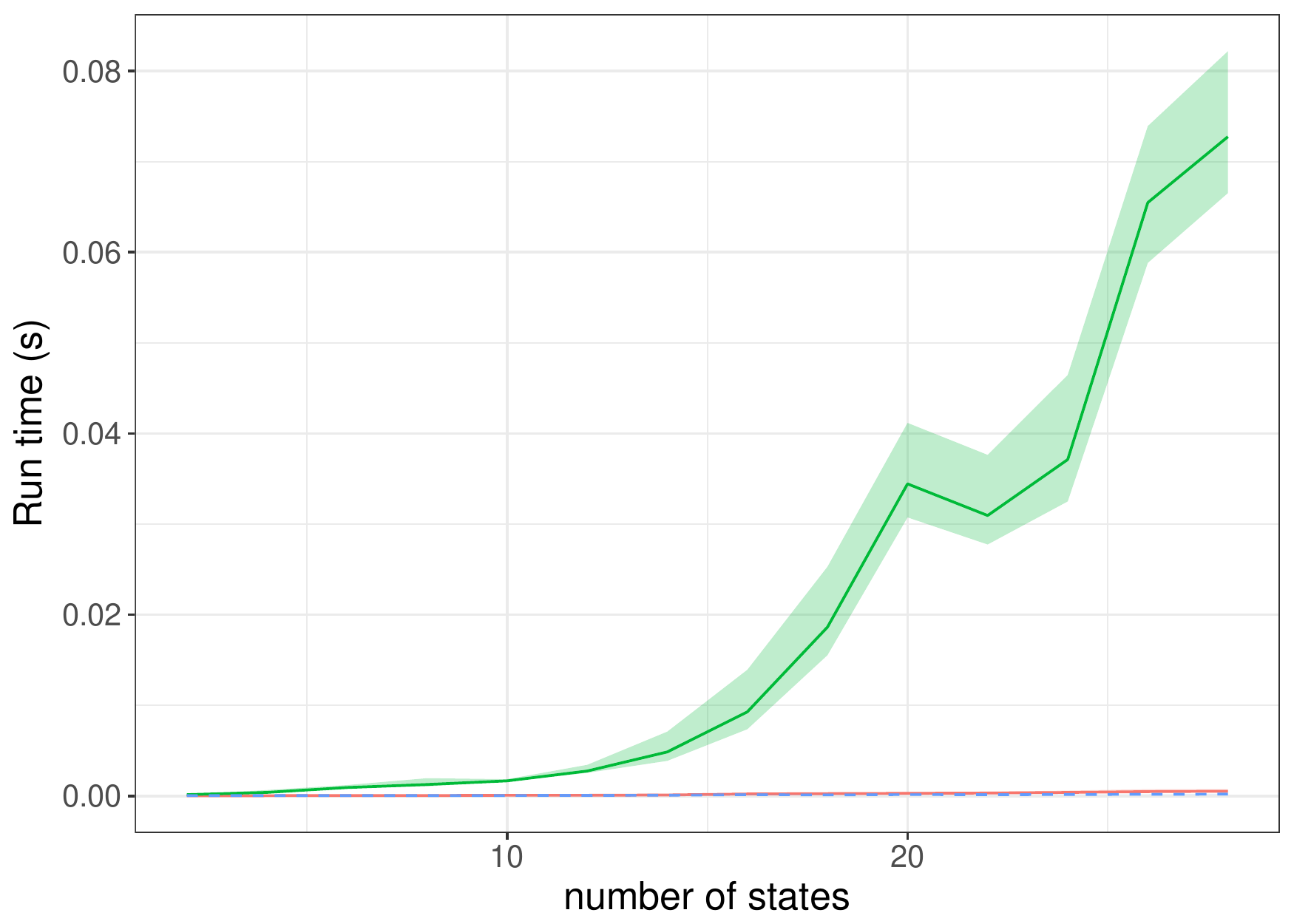}
    \includegraphics[width=8cm, height = 7cm]{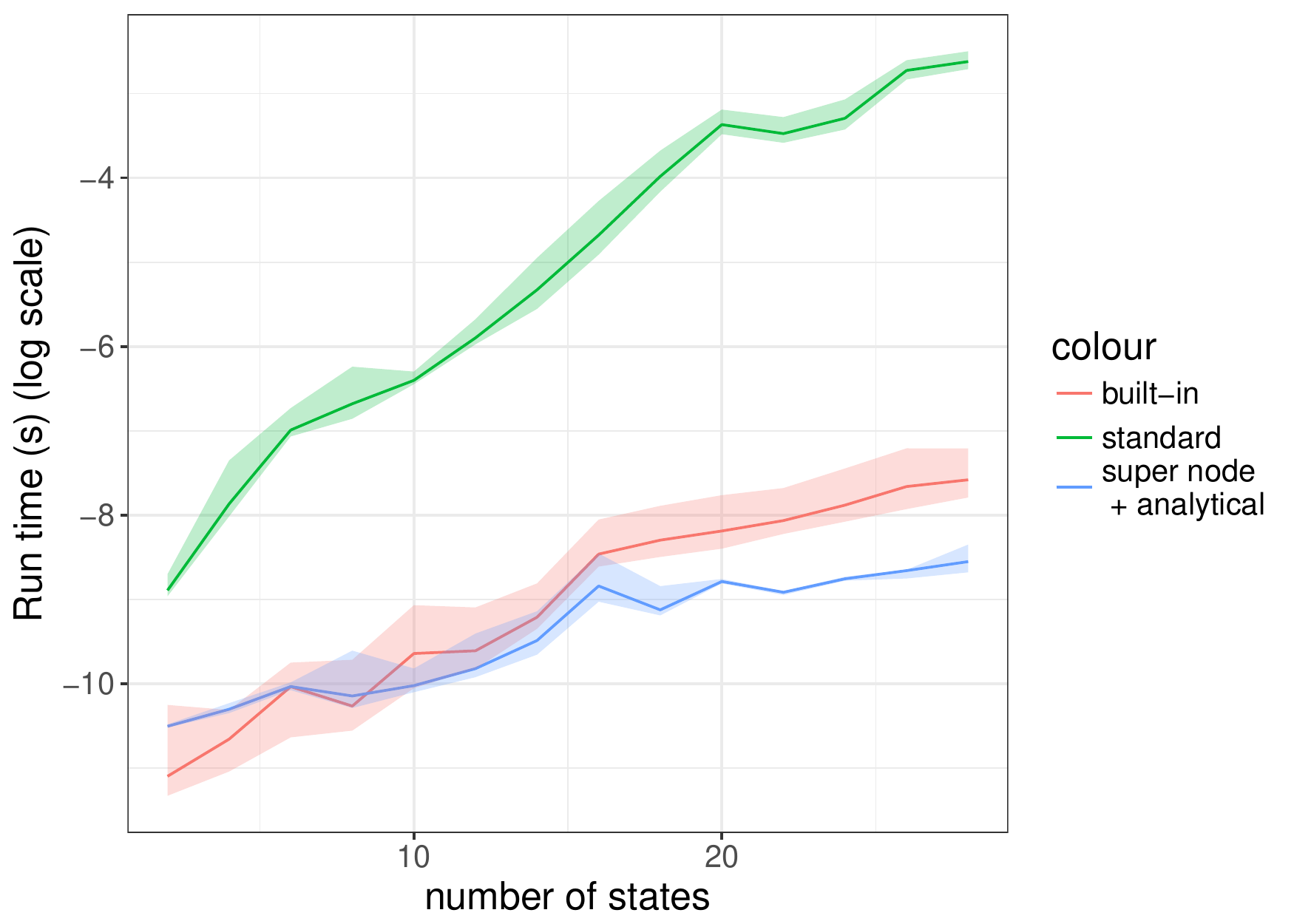}
    \caption{Run time to solve and differentiate a system of algebraic equations.
      \textit{All three solvers deploy Powell's dogleg method, an iterative algorithm that uses gradients to find the root, $y^*$, 
      of a function $f = f(y, \theta)$.
      The standard solver (green) uses an analytical expression for $J^y$ and AD to differentiate the iterative algorithm.
      The super node solver (blue) also uses an analytical expression for $J^y$ and the implicit function theorem to compute derivatives.
      The built-in solver (red) uses AD to compute $J^y$ and the implicit function theorem.
      The computer experiment is run 100 times and the shaded areas represent the region encompassed by the $5^\mathrm{th}$ and $95^\mathrm{th}$ quantiles.
      The plot on the right provides the run time on a log scale for clarity purposes.
      Plot generated with \texttt{ggplot2} \protect\cite{ggplot2:2009}.
      }}
    \label{fig:algebra1}
  \end{center}
  \end{figure}

\begin{figure}
  \begin{center}
    \includegraphics[width=15cm]{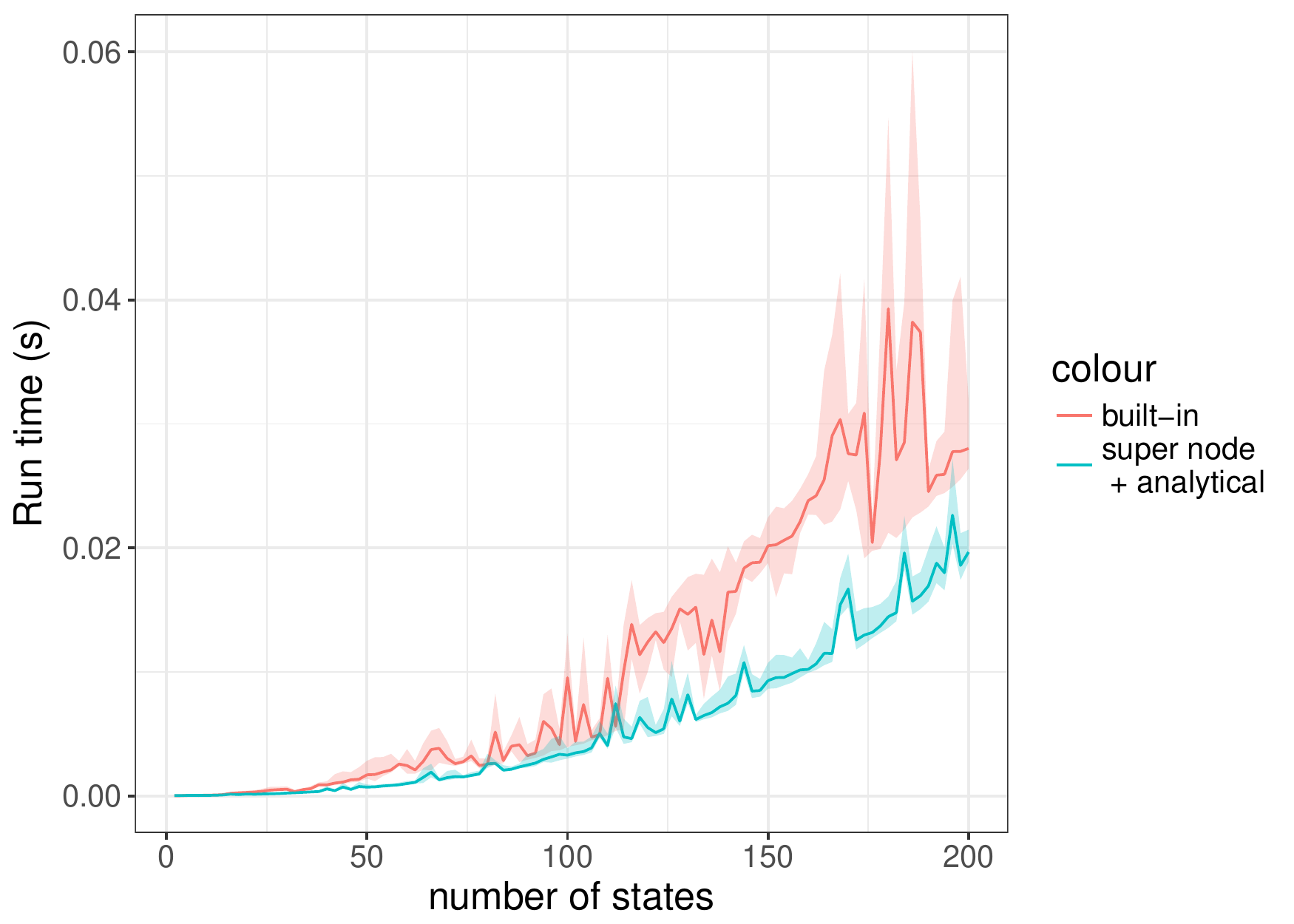}
    \caption{Run time to solve and differentiate a system of algebraic equations.
      \textit{This time we only compare the solvers which use the implicit function theorem.
      The ``super node + analytical" solver uses an analytical expression for $J^y$ both to solve the equation and compute sensitivities.
      The built-in solver uses AD to compute $J^y$. The former is approximatively two to three times faster.
      The computer experiment is run 100 times and the shaded areas represent the region encompassed by the $5^\mathrm{th}$ and $95^\mathrm{th}$ quantiles.
      Plot generated with \texttt{ggplot2} \protect\cite{ggplot2:2009}.
      }
      }
    \label{fig:algebra2}
  \end{center}
  \end{figure}

  The experiment shows using the implicit function theorem is orders of magnitudes faster than relying on standard AD
  (Figure~\ref{fig:algebra1} and table~\ref{tab:algebra1}).
  This also holds for the built-in algebraic solver which automatically differentiates $f$ with respect to $y$.
  Access to $J^y$ in its analytical form yields a minor gain in performance, which becomes more obvious when we increase the number of states
  in the algebraic equation (Figure~\ref{fig:algebra2}). 

  % natheight included for tex compiler on WIRES website

  \begin{table}
  \begin{center}
    \begin{tabular}{c c c c}
    \rowcolor[gray]{0.95} number of &  & super node & \\
    \rowcolor[gray]{0.95} states & standard & + analytical & built-in \\
    % \hline
    4 & $0.383 \pm 0.0988$ & $0.0335 \pm 0.00214$ & $0.0235 \pm 0.00842$ \\
    12 & $2.74 \pm 0.396$ & $0.0542 \pm 0.0123$ & $0.0670 \pm 0.0259$ \\
    20 & $34.4 \pm 3.28$  & $0.152 \pm 0.0145$ & $0.278 \pm 0.173$ \\
    28 & $72.8 \pm 5.56$ & $0.193 \pm 0.0301$ & $0.51 \pm 0.214$ \\
    % 50 & $550 \pm 32.0$ & $0.711 \pm 0.139$ & $1.70 \pm 0.488$\\
    % \hline
    \end{tabular}
    \caption{Run time (ms) to solve and differentiate a system of algebraic equations.
    \textit{This table complements Figure~\ref{fig:algebra1}, and highlights the results for a selected number of states.
    The uncertainty is computed by running the experiment 100 times and computing the sample standard deviation.}
    }
    \label{tab:algebra1}
  \end{center}
  \end{table}
  
  \section*{Open and practical problems}
  
  Many AD packages are still being actively developed; 
  hence we expect some of the issues we present to be addressed in the years to come.
  
  \subsection*{General and specialized code}
  
  The first problem we note is that no single package comprehensively implements all the optimization techniques outlined in this paper
  -- perhaps for several good reasons.
  First of all, two methods may not be compatible with one another.
  In most cases however, there is simply a trade-off between development effort and impact.
  This is a tricky question for strategies that are not generalizable, but  very effective at solving a specific class of problems.
  For example, retaping is not helpful when the expression graph changes from point to point and we need to evaluate the gradient at several points;
  but it is well suited for computing a single Jacobian matrix.
 
  A similar argument applies to specialized math functions.
  For some problems, such as solving ODEs and algebraic equations, we can take advantage of custom differentiation methods,
  which are agnostic to which solver gets used.
  This does not hold in general.
  Counter-examples arise when we solve partial differential equations or work out analytical solutions to field-specific problems.
  The coding burden then becomes more severe, and developers must work out which routines are worth implementing.
  % The answer should be largely user and application driven. 
  
  \subsection*{Extensibility}

  Making a package \textit{extensible} aims to reduce the burden on the developers
  and give users the coding flexibility to create new features themselves.
  Open-source projects greatly facilitate this process.
  As an example, \texttt{Torsten}\footnote{The package is still under development.
  See \url{https://github.com/metrumresearchgroup/example-models}.} extends \texttt{Stan Math} for applications in pharmacometrics \cite{Torsten:2016}.
  In our experience however, extending a library requires a great deal of time, coding expertise, and detailed knowledge of a specific package.
%  \cite{Vossback:2008} critic how obscure code for operator overloading can be;
%  we find clever schemes to manage memory, while very effective, further hurt the code's readability.
  The problem is worse when the AD package works as a black box.

  One solution is to allow users to specify custom differentiation methods when declaring a function.
  Naturally users will have to be cautioned against spending excessive time coding derivatives for minor performance gains.
%  To our knowledge, few packages provide such a feature in a high-level language.
%  \texttt{PyTorch} provides this functionality for reverse-mode AD (see \url{https://pytorch.org/docs/stable/notes/extending.html}),
%  and
  \texttt{PyTorch} provides such a feature in the high-level language \texttt{Python},
  namely for reverse-mode AD\footnote{See \url{https://pytorch.org/docs/stable/notes/extending.html}.}.
  \texttt{CasADi} \cite{Andersson:2018} 
  also offers the option to write custom derivative functions in all its interfaces
  (including \texttt{Python} and \texttt{Matlab}), though this feature is described as ``experimental" and still requires a technical
  understanding of the package and its underlying structure in
  \texttt{C++}\footnote{See \url{http://casadi.sourceforge.net/api/html/}.}.
  Another solution, which is well underway, is to educate users as to the inner workings of AD packages.
  This is in large part achieved by papers on specific packages, review articles such as this one, and well documented code.

  \subsection*{Higher-order differentiation}

  Algorithms that require higher-order differentiation are less common, but there are a few noteworthy examples.
  Riemannian Hamiltonian Monte Carlo (RHMC) is a Markov chain Monte Carlo sampler
  that requires the second and third order derivatives of the log posterior distribution,
  and can overcome severe geometrical pathologies first-order sampling methods cannot 
  \cite{Girolami:2013, Betancourt:2013}.  
  A more classical example is Newton's method, which requires a Hessian vector product when used for optimization.
  % More generally, we may be interested in computing the sensitivities of a target function which is itself a partial derivative.
  Much work has been done to make the computation of higher-order derivatives efficient.
  In general, this task remains significantly more difficult than computing first-order derivatives and can be prohibitively expensive,
  which is why we label it as a ``practical problem".
  Practitioners are indeed often reluctant to deploy computational tools that require higher-order derivatives.
  One manifestation of this issue is that RHMC is virtually never used, whereas its first-order counter part has become a
  popular tool for statistical modeling.
  
  As previously touched upon, most packages compute second-order derivatives by applying AD twice:
  first to the target function, thereby producing first-order derivatives;
  then by sorting the operations which produced these first-order derivatives into a new expression graph, and applying AD again.
  Often times, as when computing a Newton step, we are not interested in the entire Hessian matrix,
  but rather in a Hessian vector product.
 
  For simplicity, consider the scalar function $f: \mathbb R^n \to \mathbb R$.  
  To compute a Hessian vector product, one can first apply a forward sweep followed by a reverse sweep \cite{Pearlmutter:1994, Griewank:2008}.
  For second-order derivatives, AD produces an object of the general form
  \begin{eqnarray*}
    u \nabla f + w \nabla^2 f \cdot v
  \end{eqnarray*}
  where $\nabla f$ designates the gradient vector and $\nabla^2 f$ the Hessian matrix,
  and where  $v \in \mathbb R^n$, $u \in \mathbb R$, and $w \in \mathbb R$ are initialization vectors (possibly of length 1).
  The Hessian vector product $\nabla^2 f \cdot \nu$  is then obtained by setting $u = 0$, $w = 1$, and $v = \nu$,
  an efficient operation whose complexity is linear in the complexity of $f$.
  The full $n \times n$ Hessian matrix is computed by repeating this operation $n$ times and using basis vectors for $v$.
  As when computing Jacobian matrices, the same expression graph can be stored in a tape and reused multiple times.
  It is also possible to exploit the structural properties of the Hessian as is for example done by \cite{Gebremedhin:2009}
  with the package \texttt{Adol-C}.
%  For example, \cite{Gebremedhin:2009} couple AD and \textit{coloring} to efficiently compute Hessians; they demonstrate
%  their approach with \texttt{Adol-C}, a package that can detect matrix sparsity.
 
  Consider now the more general case, where we wish to compute higher-order derivatives for a function $f: \mathbb R^n \to \mathbb R^m$. 
  We can extend the above procedure: apply AD, sort the AD operations into a new expression graph, and repeat until we get the desired 
  differentiation order.

  Whether or not the above approach is optimal is an open question, most recently posed by \cite{Nomad, Betancourt:2018},
  who argues recursively applying AD leads to inefficient and sometimes numerically unstable code.
  There exists two alternatives to compute higher-order derivatives. We only briefly mention them here
  and invite the interested reader to consult the original references.
  The first alternative uses univariate Taylor series and begins with the observation that the $i^\mathrm{th}$ coefficient of the series 
  corresponds to the $i^\mathrm{th}$ derivative of our target function; more is discussed in \cite{Bischof:1993, Griewank:2000}.
  Particular attention is given to the computation of Hessian matrices and ways to exploit their structure.
%  \footnote{There
%  is in fact extensive literature on exploiting the structure of Hessian matrices to perform efficient AD.
%  See for example \cite{Griewank:2008, Gebremedhin:2009, Gower:2011}.}.
  
  For the second alternative, \cite{Betancourt:2018} proposes a theoretical framework
  for higher-order differential operators and derives these operators explicitly for the second and third oder cases.
  In some sense, this approach imitates the scheme used for first-order derivatives:
  create a library of mathematical operations for which higher-order partial derivatives are analytically worked out
  and compute sensitivities using ``higher-order chain rules''.
  Doing so is a tricky but doable task that requires a good handle on differential geometry.
  The here described strategy is prototyped in the \texttt{C++} package \texttt{Nomad} \cite{Nomad},
  which serves as a proof of concept.
  At this point however \texttt{Nomad} is not optimized to compute first-order derivatives
   and more generally needs to be further developed in order to be extensively
   used\footnote{Personal communication with Michael Betancourt.}.

  \section*{\sffamily \Large Conclusion}
  
  In recent years, AD has been successfully applied to several areas of computational statistics and machine learning.
  One notable example is Hamiltonian Monte Carlo sampling, in particular its adaptive variant the No-U-Turn sampler (NUTS) \cite{Hoffman:2014}.
  NUTS, supported by AD, indeed provides the algorithmic foundation for the probabilistic languages \texttt{Stan} \cite{Stan:2017}
  and \texttt{PyMC3} \cite{PyMC3:2016}.
  AD also plays a critical part in \textit{automatic differentiation variational inference} \cite{Kucukelbir:2016},
  a gradient based method to approximate Bayesian posteriors.
  Another major application of AD is neural networks, as done in \texttt{TensorFlow} \cite{TensorFlow:2016}.
  
  We expect that in most cases AD acts as a black box supporting a statistical or a modeling software;
  nevertheless, a basic understanding of AD can help users optimize their code,
  notably by exploiting some of the mathematical techniques we discussed.
  Advanced users can edit the source code of a package, 
  for example to incorporate an expression templates routine or write a custom derivative method for a function.

  Sometimes, a user must pick one of the many tools available to perform AD.
  A first selection criterion may be the programming language in which a tool is available;
  this constraint is however somewhat relaxed by packages that link higher-level to lower-level languages,
  such as \texttt{RCpp} which allows coders to use \texttt{C++} inside \texttt{R} \cite{Rcpp}. 
  Passed this, our recommendation is to first look at the applications which motivated the development of a package
  and see if these align with the user's goals.
  \texttt{Stan Math}, for instance, is developed to compute the gradient of a log-posterior distribution, specified by a probabilistic model.
  To do so, it offers an optimized and flexible implementation of reverse AD,
  and an extensive library of mathematical functions to do linear algebra, numerical calculations, and compute probability densities.
  On the other hand, its forward-mode is not as optimized and well tested as its reverse-mode AD\footnote{Personal communication with Bob Carpenter.}.
  More generally, we can identify computational techniques compatible with an application of interest
  (see Table~\ref{Tab:CompTechniques}),
  and consider packages that implement these techniques.
  Another key criterion is the library of functions a package provides: using built-in functions saves time and effort,
  and often guarantees a certain degree of optimality.
  For more esoteric problems, extensibility may be of the essence.
  Other considerations, which we have not discussed here, include a package's capacity to handle
  more advanced computational features, such as parallelization and GPUs.

  \section*{\sffamily \Large Acknowledgments}
    I thank Michael Betancourt, Bob Carpenter, and Andrew Gelman for helpful comments and discussions.
  
  \section*{Appendix: computer experiments}
  
  This appendix complements the section on \textit{Mathematical implementations} 
  and describes in more details the presented performance tests.   
  The two computer experiments are conducted using the following systems:
  \begin{itemize}
    \item Hardware: Macbook Pro computer (Retina, early 2015) with a 2.7 GHz Intel Core i5 with 8 GB of 1867 MHz DDR3 memory
    \item Compiler: clang++ version 4.2.1
    \item Libraries: \texttt{Eigen} 3.3.3, \texttt{Boost} 1.66.0, \texttt{Stan Math} 2.17 - develop\footnote{Versioned branch \texttt{test/autodiff\_review}.}
  \end{itemize}
  The compiler and libraries specifications are those used when runing unit tests in \texttt{Stan Math}.
  We measure runtime by looking at the wall clock before and after evaluating a target function $f$ and calculating all the sensitivities of interest.
  The wall time is computed using the \texttt{C++} function \texttt{std::chrono::system\_clock::now()}. The code is available on GitHub.
  
  \subsection*{Differentiating the $2 \times 2$ Matrix Exponential}
  
  We obtain $2 \times 2$ matrices by generating four elements from a uniform distribution $U(1, 10)$,
  that is a uniform with lower bound 1 and upper bound 10.
  This conservative approach insures the term $\Delta$ in Equation~\ref{Eq:2x2MatrixExp} is real (i.e. has no imaginary part).
  We generate the random numbers using \texttt{std::uniform\_real\_distribution<T> unif(1, 10)}.
  
  \subsection*{Differentiating a numerical algebraic solver}
  
  To test the performance of an algebraic solver, we consider a standard problem in pharmacometrics,
  namely the computation of steady states for a patient undergoing a medical treatment.
  We only provide an overview of the scientific problem,
  and invite the interested reader to consult \cite{Margossian:2018} or any standard textbook on pharmacokinetics.

  In our example, a patient orally receives a drug which diffuses in their body and gets cleared over time.
  In particular, we consider the \textit{one compartment model with a first-order absorption from the gut}.
  The following ODE system then describes the drug diffusion process:
  \begin{eqnarray*}
  \begin{aligned}
  \frac{\mathrm d y_1}{\mathrm d t} & =  - k_1 y_1 \\
  \frac{\mathrm d y_2}{\mathrm d t} & = k_1 y_1 - k_2 y_2
  \end{aligned}
  \end{eqnarray*}
  where
  \begin{itemize}
    \item $y_1$ is the drug mass in the gut
    \item $y_2$ the drug mass in a central compartment (often times organs and circulatory systems,
             such as the blood, into which the drug diffuses rapidly)
    \item $k_1$ and $k_2$ represent diffusion rates
    \item and $t$ is time.
 \end{itemize}
 This system can be solved analytically. 
 Given an initial condition $y_0$ and a time interval $\delta t$, 
 we can then define an \textit{evolution operator} $g(y_0, \delta t)$ which evolves the state of a patient, as prescribed by the above ODEs.
 
  Often times, a medical treatment undergoes a cycle:
  for example, a patient recieves a drug dose at a regular time interval, $\delta t$.
  We may then expect that, after several cycles, the patient reaches a state of equilibrium.
  Let $y_{0^-} = \{y_1(0), y_2(0)\}$ be the drug mass at the beginning of a cycle,
  $m$ the drug mass (instantaneously) introduced in the gut by a drug intake,
  and $y_{0^+} = \{y_1(0) + m, y_2(0)\}$. Then equilibrium is reached when:
  \begin{eqnarray*}
    f(y_{0^-}) = g(y_{0^+}, \delta t) - y_{0^-} = 0
  \end{eqnarray*}
  This algebraic equation can be solved analytically, but for demonstrating purposes, we solve it numerically.
  We then compute sensitivities for $k_1$ and $k_2$.
  
  To increase the number of states, we solve the algebraic equations for $n$ patients, yielding 2$n$ states.
  The coefficients in the ODEs for the $i^\mathrm{th}$ patient are $k_i = \{k_{1 i}, k_{2 i}\}$ .
  These coefficients vary from patient to patient, 
  according to $k_i =  \phi_i k$, where $k$ is a population parameter and $\phi_i$ a random scaling factor,
  uniformly sampled between 0.7 and 1.3.
  Hence, for $2n$ states, we have $2n + 2$ parameters that require sensitivities.

\bibliography{custom.bib}

\end{document}